\documentclass[aps,pra,preprint,showpacs,floatfix]{revtex4}
\usepackage{bm}
\usepackage{graphicx} 
\usepackage{amsmath}
\usepackage{color}
\usepackage{subfigure}
\usepackage{epsfig}
\begin{document}
\title{Quantum Defect Theory for cold chemistry with product quantum state resolution}
\author{Jisha Hazra$^1$, Brandon P. Ruzic$^2$, John L. Bohn$^2$, and N. Balakrishnan$^1$ }
\affiliation{$^1$Department of Chemistry, University of Nevada Las Vegas, Las Vegas,
  Nevada 89154}
\affiliation{$^2$JILA, NIST, and Department of Physics, University of Colorado, Boulder, Colorado 80309, USA}

\date{\today}

\begin{abstract}

We present a formalism for cold and ultracold atom-diatom chemical reactions that combines a quantum close-coupling 
method at short-range with quantum defect theory at long-range.  The method yields full state-to-state 
rovibrationally resolved cross sections as in standard close-coupling (CC) calculations but at a considerably less
computational expense.
This hybrid approach exploits the simplicity  of MQDT while treating the short-range interaction 
explicitly using quantum CC calculations. The method, demonstrated for D+H$_2\to$ HD+H collisions with  rovibrational quantum 
state resolution of the HD product, is shown to be accurate for a wide range of collision energies and 
initial conditions. The hybrid CC-MQDT 
formalism may provide an alternative approach 
to full CC calculations for cold and ultracold reactions.
\end{abstract}


\maketitle

\section{Introduction}

As a potentially sensitive probe of chemical reaction dynamics, ultracold molecules show great promise.
In gaseous molecular samples whose 
temperature drops well below the milliKelvin range, collision cross sections are no longer thermally averaged over a range of impact parameters 
(more properly, partial waves).  This circumstance raises the possibility of preparing reactants in individual quantum 
states of relative motion as well as internal states.  Ultracold molecules can moreover be manipulated using electric, 
magnetic, optical, and microwave fields, 
extending the reach of state preparation and therefore in principle the detail with which collision experiments can probe 
reactions~\cite{roman-review,Weck,Hutson1,Hutson2,Krems2008}.

This kind of control has been demonstrated in a prototype experiment, involving KRb molecules~\cite{Ni2008,Ospelkaus}.  
In this case reaction rates were tuned by orders of magnitude by exploiting quantum statistics of the molecules, 
the influence of electric fields, and confinement to optical lattices.  In all cases, the key ingredient to controlling the kinematics 
of the reactants lay in the manipulation of long-range forces between them.  This aspect of control distinguishes long-range physics, 
where control is applied, from short-range physics, where the chemistry actually occurs. 
For this reason, theoretical approaches that seek to understand controlled ultracold chemistry would do well to address the short- and 
long-range physics separately, yet be able to weld them together. 

While quantum close coupling (CC) calculations can handle this disparity in energy scales in principle, the computations become
impractically large
when all the relevant quantum states and external field effects are included. As a result, the vast majority
of CC calculations of ultracold reactions (both barrier and barrierless) reported so far have been 
restricted to field-free cases without the inclusion of spin and hyperfine splitting~
\cite{roman-review,Weck,Hutson1,Hutson2,Krems2008,bala-cpl-2001,Soldan02,Quemener04,Cvitas05b,Juan11,Gagan13}. Though the 
theory of chemical reactions for an atom-diatom system in external fields has been formulated by
Tscherbul and Krems~\cite{Krems} and applied to the Li+HF/LiF+H reaction, the computations remain demanding.

An alternative and simplified description, useful at ultralow temperatures,  is provided by the Multichannel Quantum Defect Theory (MQDT). 
The  formalism  was originally developed by Seaton and Fano~\cite{Seaton1,Seaton2,Fano} to understand the spectra of Rydberg atoms. Since then,
it has been successfully extended to more general contexts~\cite{Mies1,Greene84_PRA} and applied to 
both resonant and non-resonant scattering in a variety of atomic collision processes~\cite{Mies2,Burke,Raou,Gao1,Hanna09_PRA,Ruzic13_PRA}, 
ion-atom collisions~\cite{Id,Gao2},
atom-molecule systems~\cite{Hud1,Hud2}, and molecule-molecule reactive scattering~\cite{Id1,Id2,Gao3,Wang}. The method has proven flexible 
and adept at handling ultralow collision energies and field dependencies, while adequately and simply treating also the short-range physics, 
which occurs on far greater energy scales. The MQDT method has been applied to estimate overall reaction rate coefficients of several
barrierless reactions~\cite{Id1,Gao3} at far less computational expense in comparison to full CC calculations. However, in its current 
implementation, the method 
is restricted to estimating the total reaction rate coefficient. Rotational and vibrational populations of 
the reaction products are not available from these calculations, as these degrees of freedom were not included.

The value of MQDT in ultracold chemistry calculations arises from exploiting the vast disparity between the relevant energy scales.
In the cold gas, translational kinetic energies of the reactants are on the scale of milliKelvin or less.  This energy scale must necessarily
be resolved, necessitating a fine energy grid. Moreover, applied electric and magnetic fields may influence scattering on this scale, notably 
by shifting narrow resonances.  At the same time, these energy scales are dominant at large interparticle separation between the reactants.  
These circumstances play directly to the strengths of MQDT, namely, that a complete description of long-range dynamics can be handled by 
treating the channels independently, employing the Jacobi coordinates between reactants.  Doing so can expedite the calculation tremendously,
enabling rapid exploration of the sensitively varying energy and field dependence of scattering observables.
 
By contrast, the reaction itself takes place on, and is calculated using, relatively deep potential energy surfaces (PESs) of order 10-10,000 K.  
Thus over the milliKelvin energy range relevant to ultracold chemistry, wave functions of the reaction itself, restricted to the collision 
complex where all participating atoms are close together, depend only weakly on energy.  In this circumstance the reaction can be handled in
complete detail on a far coarser energy grid than required for the final observables.  Then the usual techniques of solving the reaction 
dynamics in hyperspherical coordinates, while computationally heavy, need to be performed at only a handful of energies
(perhaps only one energy, in favorable circumstances).
 
Making use of this disparity in energy and length scales necessitates a novel procedure for connecting the short-range physics,
described in hyperspherical coordinates \cite{Delves,Pack,Launay1,Launay2}, with the long-range physics, described in Jacobi coordinates 
and exploiting MQDT.  Describing and evaluating this connection procedure is a main goal of the present paper.   Note that in reactive channels, 
in which large translational energy ($\gg 1$mK) is released, the long-range wave functions are again indifferent to the mK energy scale of
reactants.  For this reason, the full MQDT theory need be applied only in the ultracold reactant channels, where the energy sensitivity resides.

In a previous paper~\cite{Jisha14a}, we described a hybrid close-coupling MQDT approach 
for non-reactive scattering in molecule-molecule collisions. Here, we extend the formalism to reactive scattering. 
The 
paper is organized as follows. In Section \ref{section:theory} we describe the theoretical 
formalism.
A brief review of the hyperspherical approach for reactive scattering is presented first to introduce the key terminologies and 
quantities necessary to describe reactive scattering. For ease of implementation we use the 
hyperspherical approach implemented in the ABC reactive scattering code~\cite{ABC-code}. This also enables the 
interested users to easily implement the formalism into the ABC code as it is widely used and publicly available. 
The MQDT formalism is described in Section \ref{section:theory}B.
 In Section \ref{section:d+h2} we discuss the numerical implementation of the approach to the benchmark D+H$_2(v,j)\to$ HD($v',j')$+H 
reaction with full 
resolution of the HD product rovibrational quantum states. Conclusions are presented in 
Section \ref{section:conclusions}.


\section{Theory}
\label{section:theory}

\subsection{Coupled-channel formulation of reactive scattering} 
We provide a brief review of the CC formalism for reactive scattering within the 
Delves hyperspherical coordinate (DC) system~\cite{Delves}. Our discussion of the reactive scattering formalism follows closely the 
description given by Pack and Parker~\cite{Pack} and Tscherbul and Krems~\cite{Krems}.

The hyperspherical coordinates for three-particle systems involve three internal coordinates (hyperradius $\rho$ and two hyperangles) 
and three external coordinates (three Euler angles). In an atom-diatom system, such as 
A+BC, there are
three arrangement channels corresponding to A+BC, AB+C and AC+B atom-diatom combinations. These three 
arrangement channels are denoted by the index $\tau$ in our notation.
The hyperradius $\rho$ and hyperangle $\theta_{\tau}$ in DC can be written in terms of two mass-scaled Jacobi distances
($S_{\tau},s_{\tau}$)
through a polar transformation: 
\begin{eqnarray}
 \rho = (S_{\tau}^2+s_{\tau}^2)^{1/2}\\
 \theta_{\tau} = \tan^{-1}\left[\frac{s_{\tau}}{S_{\tau}}\right],
\end{eqnarray}
where $S_{\tau}$ and $s_{\tau}$ are, respectively, the atom-diatom center-of-mass distance and 
internuclear separation of the diatom for a given atom-diatom arrangement.
These two coordinates along with, $\gamma_{\tau}$, the angle between the vectors ${\bf S}_{\tau}$ and ${\bf s}_{\tau}$,
form the three internal coordinates in the DC system. The three Euler angles, 
$\alpha,~\beta,~\eta$ form the external coordinates.

The  Hamiltonian  for the atom-diatom system in DC can be expressed as \cite{Pack} 
\begin{equation}
 \hat{H} = -\frac{\hbar^2}{2\mu\rho^5}\frac{\partial}{\partial \rho}\rho^5\frac{\partial}{\partial \rho}
 +\hat{H}_{\rm ad}(\rho),
 \label{DelvesHaml}
\end{equation}
where $\mu$ is the three-body reduced mass  given by $\mu=\left[\frac{m_Am_Bm_C}{m_A+m_B+m_C}\right]^{1/2}$. 
The second  term of eq.(\ref{DelvesHaml}), $\hat{H}_{\rm ad}(\rho)$ is the adiabatic 
Hamiltonian for the surface functions and is expressed as
\begin{equation}
 \hat{H}_{\rm ad}(\rho) = \frac{{\bf \hat{\ell}}^2_{\tau}}{2\mu\rho^2\cos^2\theta_{\tau}}
 +\left[V(\rho,\theta_{\tau},\gamma_{\tau})-V_{\tau}(s_{\tau})\right]+\hat{H}_{\rm mol},
\label{H_ad}
\end{equation}
where ${\bf \hat{\ell}}_{\tau}$ is the
orbital angular momentum due to end-over-end rotation of the atom-diatom system, $V$ is the total potential energy and $V_{\tau}$ is the diatomic
interaction potential when one atom is far from the other two atoms within an arrangement. The 
last term in eq.(\ref{H_ad}), the molecular Hamiltonian for a given 
diatomic fragment is given by 
\begin{equation}
 \hat{H}_{\rm mol} = -\frac{\hbar^2}{2\mu}\frac{1}{\rho^2\sin^2 2\theta_{\tau}}\frac{\partial}{\partial\theta_{\tau}}\sin^2 2\theta_{\tau}
 \frac{\partial}{\partial\theta_{\tau}}+\frac{{\bf j}_{\tau}^2}{2\mu\rho^2\sin^2\theta_{\tau}}+V_{\tau}(s_{\tau}),
 \label{MolHaml}
\end{equation} where ${\bf j}_{\tau}$ is the rotational angular momentum of the diatomic species in the 
arrangement channel $\tau$.

An adiabatic approach is used to solve the Schr\"{o}dinger equation in hyperspherical coordinates.
This involves partitioning 
the hyperradius into a large number of sectors, and within each sector, diagonalizing the Hamiltonian ${\hat H}_{\rm ad}$ in the 
remaining degrees of freedom:
\begin{equation}
 \hat{H}_{\rm ad}(\rho) \Phi^{JM}_{n}(\omega;\rho) = \epsilon_n(\rho)\Phi^{JM}_{n}(\omega;\rho),\label{AdiabaticHam}
\end{equation}
where $n = 1, 2, \dots, N$, and $N$ is the total number of adiabatic states retained in the calculation. 
In the limit $\rho \rightarrow \infty$, each index $n$ correlates to a set $\{\tau,v,j,\ell\}$ denoting vibrational ($v$), 
 rotational ($j$),  and orbital angular momentum ($\ell$)
quantum numbers within each arrangement channel, $\tau$.
This yields the surface functions $\Phi^{JM}_{n}(\omega;\rho)$, where 
$\omega$ collectively denotes the two hyperangles (internal angles
$\theta_{\tau}$ and $\gamma_{\tau}$)
and the corresponding eigenvalues (adiabatic energies $\epsilon_n(\rho)$). 
The quantum numbers $J$ and $M$ specify the total 
angular momentum ($\vec{J}= \vec{j} + \vec{\ell}$) and its projection on a space-fixed (SF) axis. 
The $\rho$ dependence of the adiabatic energies arises from 
the parametric
dependence of the adiabatic Hamiltonian on $\rho$. 
 Note that $\hat{H}_{\rm ad}(\rho)$ is the Hamiltonian operator for the three-body system without the radial kinetic energy operator.
 
To make the computation of the surface functions numerically efficient,
  they are further expanded \cite{ABC-code} in terms of primitive orthonormal
basis sets $\xi_m(\omega;\rho)$ for a given $J$ and $M$:  
\begin{eqnarray}
\Phi^{JM}_n(\omega;\rho) &=& \sum_m F_{mn}(\rho) \xi_m (\omega;\rho),\nonumber 
 \end{eqnarray}
 where $F_{mn}(\rho)$ are the expansion coefficients and
 \begin{equation}
\xi_m(\omega;\rho) = \frac{1}{\sqrt{\sigma_m}}\frac{2}{\sin 2\theta_{\tau}}\sum_{\tau,v,j,\ell}X_{\tau vj\ell,m}(\rho)
\Upsilon_{\tau vj}(\theta_{\tau};\rho){\cal Y}^{JM}_{j\ell}({\hat s}_{\tau},{\hat S}_{\tau})\label{Primitive}.
\end{equation}
The quantities $\sigma_{m}$ and $X_{\tau vj\ell,m}$ in the above equation are the eigenvalues and eigenvectors of the overlap matrix ${\bf O}$,  
defined in Appendix-A. The functions  
${\cal Y}^{JM}_{j\ell}({\hat s}_{\tau},{\hat S}_{\tau})$ are the rotational wave functions of
the atom-diatom system in the total angular momentum representation and $\Upsilon_{\tau vj}$ are $\rho$-dependent
vibrational wave functions of the diatomic fragment in each arrangement.
Although, the vibrational functions $\Upsilon_{\tau vj}(\theta_{\tau};\rho)$ are completely 
orthonormal within a given $\tau$,  they are not orthogonal between different $\tau$. 
Due to non-zero overlap of $\Upsilon_{\tau vj}(\theta_{\tau};\rho)$ between the different 
arrangement channels  at small $\rho$, a canonical orthogonalization of the basis, as given by Eq.~(\ref{Primitive}), via the 
overlap matrix ${\bf O}$ is required to construct the appropriate orthogonal primitive basis sets  between different $\tau$ \cite{Krems}. 
  The vibrational wave functions, $\Upsilon_{\tau vj}(\theta_{\tau};\rho)$, and the corresponding eigenenergies, $\epsilon_{\tau vj}$,
 of the diatomic species are solutions of the eigenvalue problem involving the molecular Hamiltonian carried out at each value of the hyperradius:
 \begin{equation}
\left(-\frac{\hbar^2}{2\mu\rho^2}\left[\frac{\partial^2}{\partial\theta_{\tau}^2}-\frac{j_{\tau}(j_{\tau}+1)}{\sin^2\theta_{\tau}}\right]
+V_{\tau}(s_{\tau})\right)\Upsilon_{\tau vj}(\theta_{\tau};\rho)
= \epsilon_{\tau vj}(\rho) \Upsilon_{\tau vj}(\theta_{\tau};\rho).
\end{equation} 
Note that in the asymptotic limit, the adiabatic energies $\epsilon_n(\rho)$ coincide with the
eigenenergies $\epsilon_{\tau vj}$.

The adiabatic surface functions $\Phi_{n}^{JM}(\omega;\rho)$ serve as the basis functions for expanding the 
 total wave function $\Psi^{JM}$ of the triatomic system:
\begin{equation}
 \Psi^{JM}(\rho) =\frac{1}{\rho^{5/2}}\sum_{n}\Gamma^J_{n}(\rho)\Phi_{n}^{JM}(\omega;\rho), \label{Totalwf}
\end{equation}
 where $\Gamma^J_{n}(\rho)$ is a $\rho$-dependent radial solution.
On substitution  of Eq.~(\ref{Totalwf}) into the time-independent Schr\"{o}dinger equation $H\Psi^{JM}=E_{\rm tot}\Psi^{JM}$
one obtains radial equations of the form
\begin{equation}
\frac{d^2 \Gamma_{n^{\prime}}^{J}(\rho)}{d\rho^2}=\sum_n W_{n^{\prime}n}(\rho)\Gamma_{n}^{J}(\rho)\label{radialwf}
\end{equation}
with the matrix elements, 
\begin{equation}
  W_{n^{\prime}n}(\rho) = \frac{2\mu}{\hbar^2}\left[\epsilon_n(\rho)+\frac{\hbar^2}{8\mu\rho^2}-E_{tot}\right] \delta_{n^{\prime}n}+
 P_{n^{\prime}n} + Q_{n^{\prime}n},
\end{equation}
where $\bf P$ and $\bf Q$ are derivative coupling matrices that account for the action of the hyperradial kinetic energy 
$-(\hbar^2/2\mu\rho^5)\partial / \partial\rho (\rho^5 \partial / \partial \rho)$ on the $\rho$-dependent adiabatic basis functions.
Note that Eq. (\ref{radialwf}) is written for a given sector, within which the surface functions 
$\Phi^{JM}_n(\omega;\rho)$ are assumed to
be independent of $\rho$.
Therefore, the derivative coupling matrices $P$ and $Q$ are 
neglected within a sector. 
However, the adiabatic surface functions
$\Phi^{JM}_n(\omega;\rho)$ vary with $\rho$, and a ``sector adiabatic" technique is adopted for radial
integration in $\rho$. By dividing the entire range of $\rho$ 
into small sectors and enforcing continuity of the radial wavefunctions and their first derivatives at the boundary of each sector, 
the solutions $ \Gamma_{n}^{J}(\rho)$ are transformed at the boundary between the $j$th to $(j+1)$th sectors using the relationship 
\begin{equation}
{\bf Y}(\rho_{j+1}) = {\bf S}^{\rm T}(\rho_{j},\rho_{j+1}){\bf Y}(\rho_j){\bf S}(\rho_{j},\rho_{j+1}),\label{prop}
\end{equation}
where ${\bf Y}$ denotes the log-derivative matrix ${\bf Y}(\rho_{j}) = \frac{d{\bf \Gamma}(\rho_j)}{d\rho_j} 
{\bf \Gamma}(\rho_j)^{-1}$ and the sector-to-sector transformation matrix ${\bf S}(\rho_{j},\rho_{j+1})$ is defined in Eq.~(\ref{sectoroverlap}) of
Appendix-A.
The log-derivative matrix is propagated using the diagonal reference potential method 
of Manolopoulos~\cite{Mano}.

Thus, in the present approach, the reactive scattering problem can be divided into two major steps: 
(i) solving the eigenvalue problem of Eq.~(\ref{AdiabaticHam}) 
to evaluate the surface functions and adiabatic energies and (ii) propagating the radial equations from a small $\rho$ within the 
classically forbidden region to a large asymptotic value using Eq.~(\ref{prop}). 
The first step involves (a) the construction of the overlap matrix ${\bf O}$ to evaluate the eigenvectors 
$X$ and eigenvalues $\sigma$, (b) the evaluation of the matrix elements $\langle \xi_{m}|H_{\rm ad}(\rho)|\xi_{m'}\rangle $ in the primitive 
orthogonal basis sets, and (c) the diagonalization of the above matrix to yield the adiabatic eigenvalues $\epsilon_n(\rho)$ and the 
corresponding expansion coefficients $F_{mn}(\rho)$. The expression for the matrix elements of the adiabatic
Hamiltonian is given by Eq.~(\ref{MatAdia}) in Appendix-A.
Once  $\epsilon_n(\rho)$ and $F_{mn}(\rho)$ are evaluated, in the second step, the radial Eqs.~(\ref{radialwf}) are
propagated from $\rho_{\rm min}$ to $\rho_{\infty}$ via Eq. (\ref{prop}) followed by applying scattering  boundary 
conditions to evaluate the reactant matrix
${\bf K}^J$ and scattering matrix ${\bf S}^J$. Thus far, everything is formulated in the SF coordinates,
and one can  
propagate the radial equations in this coordinate system. However, Eqs.~(\ref{overlapmat}), (\ref{MatAdia}), and (\ref{sectoroverlap}) 
of Appendix-A
involve five-dimensional integrals that are hard to evaluate and 
computationally intractable. This difficulty can be overcome by transforming the 
angular functions ${\cal Y}^{JM}_{j\ell}({\hat s}_{\tau},{\hat S}_{\tau})$ in these equations from
SF to the body-fixed (BF) frame. This entails the transformation of the radial 
wave function from SF hyperspherical coordinates to BF hyperspherical co-ordinates.
Thus, for computational efficiency, the radial wave functions $\Gamma_{n'}^{J}(\rho)$ are propagated according to Eq.(\ref{radialwf})
in the BF representation. See Refs. \cite{Pack,Krems} for details of the SF to BF transformation. Details of 
the asymptotic matching procedure are 
described in Appendix-B.

In the ABC code \cite{ABC-code}, before applying asymptotic boundary conditions, the log-derivative matrix at 
the last sector in $\rho$, is transformed 
from the BF to the SF representation in Delves hyperspherical coordinates. Asymptotic boundary conditions
are then applied to the log-derivative matrix in SF coordinates (as described in Appendix-B) to evaluate 
the reactance and the scattering matrices. The scattering matrix is subsequently transformed from  SF  to BF representation to compute 
the standard helicity-representation
S-matrix~\cite{ABC-code}. This is because the ABC code is formulated in BF coordinates.

\subsection{CC-MQDT Approach for Matching to Asymptotic Wave Functions}

Having constructed the log-derivative matrix in hyperspherical coordinates, 
the scattering calculation next needs to continue the solution to asymptotically 
large values of the relative coordinate $S_{\tau}$ of the reactants or products. 
In any form of scattering theory, this is accomplished by using ${\bf Y}$ as a boundary
condition to construct linear combinations of asymptotic wave functions in the coordinate $S_{\tau}$,
for values of $S_{\tau}$ larger than a convenient matching distance $S_{m}$.  It is assumed that the 
scattering channels are uncoupled for $S_{\tau} \ge S_{m}$, whereby the complete wave function is a linear 
combination of solutions in each channel separately.  This linear combination is conventionally given as
\begin{equation}
\label{definition_of_K}
 M_{fi}(S_{\tau}) =  \hat{f}_f(S_{\tau})\delta_{fi}- \hat{g}_f(S_{\tau}) K^{\rm sr}_{fi}, \;\;\;\; S_{\tau}\ge S_m.
\end{equation}
Here, $\hat{f}_i$ and $\hat{g}_i$ represent a pair of linearly independent reference functions in each channel, 
satisfying a Schr\"{o}dinger equation
\begin{eqnarray}
\left( -\frac{ \hbar^2 }{ 2\mu  } \frac{ d^2 }{ dS_{\tau}^2 }  + \frac{ \hbar^2 \ell_i (\ell_i + 1) }{ 2 \mu S_{\tau}^2 } 
+V^\text{lr}(S_{\tau}) \right) \left\{ \begin{array}{c} \hat{f}_i \\ \hat{g}_i \end{array} \right\} = E_{c_i} 
\left\{ \begin{array}{c} \hat{f}_i \\ \hat{g}_i \end{array} \right\}\label{1D},
\end{eqnarray}
where $\mu$ is the reduced mass of three-body system as defined earlier,
$\ell_i$ is their relative partial wave and $E_{c_i}$ is the kinetic energy in  channel $i$.
The channel index
$i$ asymptotically correlates with separated molecule quantum numbers $\{\tau,v,j,\ell\}$. $V^\text{lr}$ is 
the reference potential in the long-range of the form $V^\text{lr} = -\frac{C_6}{S^6_{\tau}}-\frac{C_8}{S^8_{\tau}}-\frac{C_{10}}{S^{10}_{\tau}}$.
 The detailed procedure for translating the wave function in the form of the log derivative ${\bf Y}$
in hyperspherical coordinates into an asymptotic function (\ref{definition_of_K}) in Jacobi coordinates is given in Appendix-B.  

The definition of ${\bf K}^\text{sr}$ in (\ref{definition_of_K}) is tied naturally to the definition of the reference functions. 
In the product channels, the collision energy is sufficiently large that the reference potential $V^\text{lr}$ is negligible at 
distances $S_{\tau}>S_m$, whereby the reference solutions $\hat{f}$ and $\hat{g}$ are taken as the energy-normalized
free-particle solutions $f$ and $g$.  Specifically, in open channels these are given in terms of the spherical Riccati-Bessel functions
\begin{eqnarray}
\label{ref_f}
 f_i(S_{\tau}) = k_i^{1/2} S_{\tau} j_{\ell_i}(k_iS_{\tau})\\
\label{ref_g}
 g_i(S_{\tau}) = k_i^{1/2} S_{\tau} n_{\ell_i}(k_iS_{\tau}),
\end{eqnarray}
where $k_i = \sqrt{\frac{2\mu E_{c_i}}{\hbar^2}}$. For closed channels, $f_i$ and $g_i$ are closely related to the modified spherical Bessel functions of first ($I_{\ell_i+1/2}$) and
second ($K_{\ell_i+1/2}$) kind.

By contrast, in the reactant channels where the collision energy is in the mK-$\mu$K range, the 
long-range reference potential is {\it not} negligible and must be taken into account.
As the solutions $f$ and $g$ strongly depend on energy in the threshold regime, the alternative solutions 
${\hat f}_i$ and ${\hat g}_i$ of \cite{Ruzic13_PRA} are used in the reactant channels. 
These solutions are not energy-normalized and weakly depend on energy in the threshold regime. 
Moreover, they are able to retain their linear independence in the threshold regime, even when the partial wave is nonzero.
In the present context, they are defined by the WKB-like boundary conditions \cite{Ruzic13_PRA},
\begin{eqnarray}
\label{QDT_ref_f}
\hat{f_i} (S_{\tau}) = \frac{1}{\sqrt{(k_i(S_{\tau}))}} {\rm sin} \left( \int_{S_x}^{S_{\tau}}k_i(S'_{\tau}) dS'_{\tau}+\phi_i\right)
\hspace{0.2cm} {\rm at}
\hspace{0.2cm} S_{\tau}=S_x \\
\label{QDT_ref_g}
\hat{g_i} (S_{\tau}) = -\frac{1}{\sqrt{(k_i(S_{\tau}))}} {\rm cos} \left( \int_{S_x}^{S_{\tau}}k_i(S'_{\tau}) dS'_{\tau}+\phi_i\right) 
\hspace{0.2cm} {\rm at} 
\hspace{0.2cm} S_{\tau}=S_x, 
\end{eqnarray}
at some small radius $S_x \leq S_m$, where $k_i(S_{\tau})=\sqrt{\frac{2\mu}{\hbar^2}(E_{c_i}-V^\text{lr}(S_{\tau}))}$. 
The phase $\phi_i$ is carefully chosen so as to preserve the linear independence of these functions in the asymptotic limit
\cite{Ruzic13_PRA}.
These solutions are in turn related to energy-normalized solutions via standard MQDT transformations
\begin{eqnarray}
\label{QDT_transformation}
 f_i &=& \hat{f_i}{\cal A}^{1/2}_i\\
 g_i &=& \hat{f_i}{\cal A}^{-1/2}_i{\cal G}_i+\hat{g_i}{\cal A}^{-1/2}_i,
\label{QDT_transformation2}
\end{eqnarray}
where the quantities ${\cal A}$ and ${\cal G}$ have standard forms given in Ref. \cite{Ruzic13_PRA} and crucially are smoothly dependent on energy.  

Carrying out the resulting matching procedure, using free-particle solutions 
(\ref{ref_f}) and (\ref{ref_g}) in product channels and MQDT solutions (\ref{QDT_ref_f}) and (\ref{QDT_ref_g})
in reactant channels, results in a provisional, short-range ${\bf K}$-matrix, denoted ${\bf K}^{\rm sr}$.  Its main feature in 
the theory is that it is generally only weakly dependent on energy in the ultracold regime near the reactants'
threshold, as we will show in the examples below.  Thus ${\bf K}^{\rm sr}$ can be interpolated, reducing the number of 
energies at which the full hyperspherical calculation must be performed.  At this point, the 
exponentially growing closed-channel components are eliminated, 
following the usual procedures of MQDT \cite{Ruzic13_PRA,Jisha14a} 
to yield a reduced ${\bf K}$-matrix ${\bf \tilde K}$ via the equation,
\begin{equation}
 {\bf \tilde K}={\bf K}^{\rm sr}_{oo}-{\bf K}^{\rm sr}_{oc}\left({\rm cot}\gamma+{\bf K}^{\rm sr}_{cc}\right)^{-1}{\bf K}^{\rm sr}_{co}.
\end{equation}  
This reduced ${\bf K}$-matrix is in turn conveniently rewritten in blocks corresponding to reactant and product channels, as
\[ {\bf \tilde K} = \left( \begin{array}{cc}
{\bf \tilde K}_{RR} & {\bf \tilde K}_{RP}  \\
{\bf \tilde K}_{PR} & {\bf \tilde K}_{PP} \end{array} \right).\] 
There remains only the matter of translating from the reference functions ${\hat f}_i$ and ${\hat g}_i$ into the
energy normalized versions $f_i$ and $g_i$ via eqs.(\ref{QDT_transformation})
and (\ref{QDT_transformation2}).  
In block-matrix form, this gives the transformation matrices
\[ {\bf A} = \left( \begin{array}{cc}
{ \cal A}^{1/2} & 0  \\
0 & {\bf I} \end{array} \right),\] \[ {\bf B} = \left( \begin{array}{cc}
0 & 0  \\
0 & 0 \end{array} \right),\]\[ {\bf C} = \left( \begin{array}{cc}
{\cal A}^{-1/2}{\cal G} & 0  \\
0 & 0 \end{array} \right),\]and \[ {\bf D} = \left( \begin{array}{cc}
{\cal A}^{-1/2} & 0  \\
0 & {\bf I} \end{array} \right),\]
where ${\bf I}$ is the identity matrix.
Making this substitution in the wave function (\ref{definition_of_K}) transforms ${\bf \tilde K}$ into the final $\bf K$-matrix
\begin{equation}
 {\bf K} = ({\bf D+{\tilde K}C})({\bf B+{\tilde K}A})^{-1}.  
\end{equation}
 In block notation, the final expressions for the different blocks  of the asymptotic K-matrix become
\begin{eqnarray}
 {\bf K}_{\rm RR} &=& { \cal A}^{1/2}({\bf I}+{\bf {\tilde K}}_{\rm RR}{\cal G})^{-1}{\bf {\tilde K}}_{\rm RR}{ \cal A}^{1/2}, \label{trans1}\\
 {\bf K}_{\rm PR} &=&
 {\bf {\tilde K}}_{\rm PR}\left({\bf I}-{\cal G}({\bf I}+{\bf {\tilde K}}_{\rm RR}{\cal G})^{-1}{\bf \tilde K}_{\rm RR}\right){ \cal A}^{1/2}, \label{trans2}\\
 {\bf K}_{\rm RP} &=& { \cal A}^{1/2}({\bf I}+{\bf {\tilde K}}_{\rm RR}{\cal G})^{-1}{\bf {\tilde K}}_{\rm RP}, \label{trans3}\\
 {\bf K}_{\rm PP} &=& {\bf {\tilde K}}_{\rm PP}-{\bf {\tilde K}}_{\rm PR}{\cal G}({\bf I}+{\bf {\tilde K}}_{\rm RR}{\cal G})^{-1}
 {\bf {\tilde K}}_{\rm RP} ,\label{trans4}
\end{eqnarray} The symmetry of the resulting $\bf K$-matrix
is illustrated by the following transformation:
\begin{eqnarray}
 {\bf K}_{\rm RP}^{T} &=& {\bf {\tilde K}}_{\rm PR}({\bf I}+{\cal G}{\bf {\tilde K}}_{\rm RR})^{-1}{ \cal A}^{1/2},  \\ 
 &=& {\bf {\tilde K}}_{\rm PR}\left({\bf I}-({\bf {\tilde K}}_{\rm RR}^{-1}{\cal G}^{-1}+{\bf I})^{-1}\right){ \cal A}^{1/2},\label{woddbery} \\
 &=& {\bf {\tilde K}}_{\rm PR}\left({\bf I}-{\cal G}({\bf I}+{\bf {\tilde K}}_{\rm RR}{\cal G})^{-1}{\bf \tilde K}_{\rm RR}\right){ \cal A}^{1/2}, \\
 &=& {\bf K}_{\rm PR},
 \end{eqnarray}
where we have used Woodbury's matrix identity  
to arrive at Eq. (\ref{woddbery}).  In the last step, the additional phase shift $\eta_i$ in each channel, 
due to propagation in the long-range potentials $V^\text{lr}$ in each reactant channel, must be incorporated.  
This leads to the  physical scattering matrix
\[ {\bf S}^{\rm phys} = \left( \begin{array}{cc}
e^{i\bf \eta} & 0  \\
0 & {\bf I} \end{array} \right)({\bf I}+i{\bf K})({\bf I}-i{\bf K})^{-1}\left( \begin{array}{cc}
e^{i{\bf \eta}} & 0  \\
0 & {\bf I} \end{array}\right).\]

\section{Application to D+H$_2(v,j)\to$ HD$(v',j')$ +H reaction}
\label{section:d+h2}
Since our approach is implemented in the ABC code, first we briefly summarize the key steps involved in the computation of 
reaction probabilities and cross sections using the ABC code.
For an incident kinetic energy $E_c$, 
separate runs of the ABC program are needed  for
each  value of the total angular momentum quantum number $J$, parity $p$ of the triatomic complex, and a specified value of the diatomic parity
$q$, where $q=(-1)^j$ for homonuclear diatomic molecules. 
Therefore, each triplet $\{J,p,q\}$ requires a different calculation leading to a parity-adapted scattering matrix, 
${\bf S}^{Jp}_{\tau vj\Omega \to \tau'v'j'\Omega'}$, where $\Omega$ is the projection of 
$J$ in the BF frame.
Once this matrix is evaluated, 
any observable property of the reaction can be computed by transforming ${\bf S}^{Jp}_{\tau vj\Omega \to \tau'v'j'\Omega'}$ 
into a helicity-representation ${\bf S}$-matrix, 
${\bf S}^{J}_{\tau vj\Omega \to \tau'v'j'\Omega'}$ by using appropriate expressions \cite{ABC-code}.  Hence, for a given $J$
and an incident kinetic energy $E_c$, overall quenching (non-reactive) and reaction probabilities 
from an initial quantum state 
$\{\tau, v,j,\Omega\}$ are given by
\begin{eqnarray}
 P^J_{\tau vj\Omega,{\rm qn}}(E_c) = \sum_{{\rm quenching}\hspace{0.1cm} v'j'\Omega'}
 P^J_{\tau v j\Omega \to \tau v'j'\Omega'}(E_c),\\
 P^J_{\tau v j \Omega,{\rm re}}(E_c) = \sum_{{\rm reactive}\hspace{0.1cm} v'j'\Omega'}
 P^J_{\tau v j\Omega \to \tau'v'j'\Omega'}(E_c),
\end{eqnarray}
where the state-to-state probability $P^J_{\tau vj\Omega \to \tau'v'j'\Omega'}=|S^J_{\tau vj\Omega \to \tau'v'j'\Omega'}|^2$.
The corresponding cross sections, summed over all final states, become
\begin{eqnarray}
 \sigma^{J}_{\tau vj\Omega,{\rm qn}}(E_c) =\frac{\pi}{k^2_{\tau vj\Omega}(2j+1)}
 \sum_{{\rm quenching}\hspace{0.1cm} v'j'\Omega'}|S^J_{\tau v j\Omega \to \tau v'j'\Omega'}(E_c)|^2,\\
 \sigma^{J}_{\tau vj\Omega,{\rm re}}(E_c) =\frac{\pi}{k^2_{\tau vj\Omega}(2j+1)}
 \sum_{{\rm reactive}\hspace{0.1cm} v'j'\Omega'}|S^J_{\tau v j\Omega \to \tau'v'j'\Omega'}(E_c)|^2,
\end{eqnarray}
where $k_{\tau vj\Omega}=\sqrt{\frac{2\mu E_{c}}{\hbar^2}}$ is the wave vector with respect to the initial collisional channel.
The elastic cross section is obtained as $\sigma^{J}_{\rm el}=
\frac{\pi}{k^2_{\tau vj\Omega}}|1-S^J_{\tau vj\Omega\to \tau vj\Omega}(E_{c})|^2$.
\begin{figure}[tbp]
\centering
\includegraphics[width=0.8\columnwidth]{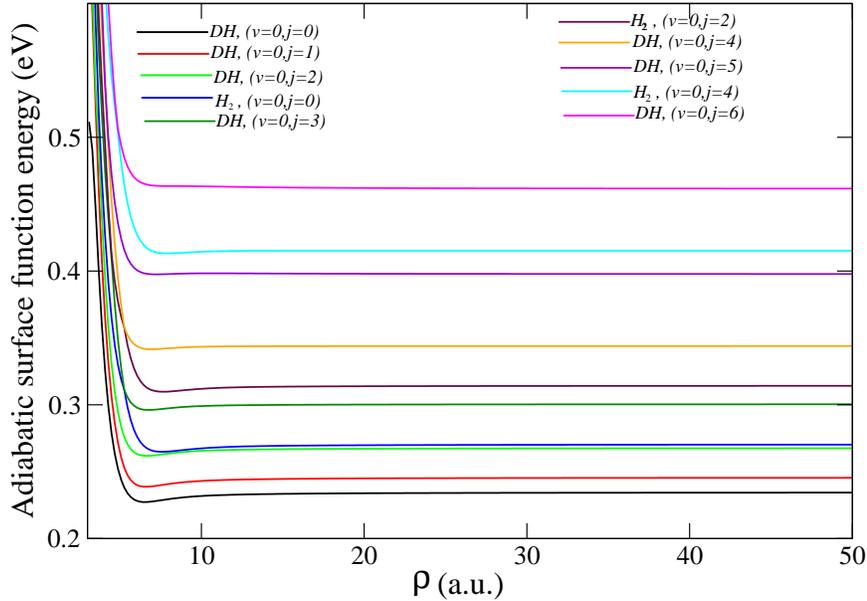}
\caption{(Color online) Lowest 10 adiabatic surface function energies $\epsilon_n(\rho)$ (as defined by Eq.(\ref{AdiabaticHam})) 
as a function of the hyperradius
$\rho$ for the D+H$_2(v=0,j=0)$ system for $J=0$. The different adiabatic curves asymptotically correspond to different
ro-vibrational levels of H$_2$ and DH molecules as indicated in the figure.}
\label{adiabatic}
\end{figure}

We implemented the above approach for reactive scattering by taking the D+H$_2(v,j)\to $ HD$(v',j')$ +H reaction
as an illustrative example. 
The method requires an accurate 
description of the long-range potential in the reactant channels, where the MQDT formalism is applied. 
Since most available PESs for elementary chemical reactions do not
provide an accurate treatment of the long-range interaction, a reliable description of reactive scattering  in ultracold
collisions continues to be a challenge. This appears to be the case, even for widely studied benchmark 
systems such as the F+H$_2$ reaction \cite{bala-cpl-2001,Tizniti14}. The choice of D+H$_2$ is motivated in part due to the availability of 
an accurate PES for this system reported by Mielke $et. ~al.$ \cite{Mielke} that includes long-range forces for the diatomic species but
also the possibility of 
doing quick tests to 
benchmark the 
results. However, due to the fairly large energy barrier for the reaction, the reactivity is 
small for rovibrationally  ground state H$_2$ molecules. Despite such low reactivity, we show that the 
MQDT formalism works very well for the ground state of H$_2$. 
Fig. \ref{adiabatic} shows the 
adiabatic surface function energies $\epsilon_n(\rho)$  for the D+H$_2$ system for $J=0$. The  different adiabatic curves asymptotically
correlate with different
ro-vibrational levels of H$_2$ and HD molecules. For the initial $v=0,j=0$ level of the H$_2$ molecule, ro-vibrational levels $v'=0,j'=0-2$
of the HD molecules are energetically accessible in the zero collision energy limit.

The MQDT reference functions and parameters
are determined by solving one-dimensional Schr\"{o}dinger equations in a reference potential of the form 
$V^{\rm lr}$ as mentioned in Eq.~(\ref{1D}).
Details are given in \cite{Ruzic13_PRA,Jisha14a}. This requires long-range expansion coeffcients
for the D+H$_2$ atom-dimer interactions.
For the diatomic fragments the  dispersion coefficients 
are accurately known. Their values in atomic units are $C_6 = 6.499027, C_8 = 124.3991$ and $C_{10} = 3285.828$ \cite{Mielke}.
However, Mielke et al. \cite{Mielke} did not report the corresponding values for the atom-diatom interaction. We numerically extracted an
effective $C_6$ by fitting the lowest diagonal element of the long-range part of the diabatic 
potential matrix to a long-range expansion of the form of $V^{\rm lr}$.
The diabatic potentials are constructed by matrix elements of the interaction potential in a basis set consisting of ro-vibrational 
wavefunctions of the
H$_2$ molecule.
The $C_6$ co-efficients evaluated this way are slightly sensitive to the initial vibrational level of H$_2$, but we have used the same 
$C_6$-coefficient obtained for $v=0$ for all 
vibrational levels investigated in this study. 
Despite this simplicity 
in the choice of the long-range potential, the hybrid CC-MQDT approach reproduces the full CC results within 5-10\% in most of the cases. 
This clearly indicates that 
the short-range reaction dynamics is fully characterized by the short-range K-matrix.

\subsection{Convergence tests}
\subsubsection{Matching Distance:}
Asymptotic boundary conditions (i.e., matching to free-particle wave functions) can be applied only when the interaction potential becomes small
compared to the collision energy. In cold and ultracold collisions this requires radial integration of the coupled equations
to large values of the hyperradius. For the full CC calculations,
converged results are obtained by matching the log-derivative matrix to free particle wave functions at 
$\rho_{\infty} = 100 ~a_0$. However, for the MQDT version, a short-range matching distance $S_m$ just outside the
region of chemical interaction is used to minimize the computational cost. In this case
the CC calculation needs to be performed only up to $S_m$, 
beyond which the whole multichannel scattering problem is converted into 
a single channel problem involving a subset of channels ($N'$) treated by MQDT.
 This leads to computational effort proportional to $N'<N$ (not $N^3$) beyond $S_m$. 

\begin{figure}[tbp]
\centering
\begin{tabular}{cc}
\includegraphics[width=0.32\columnwidth]{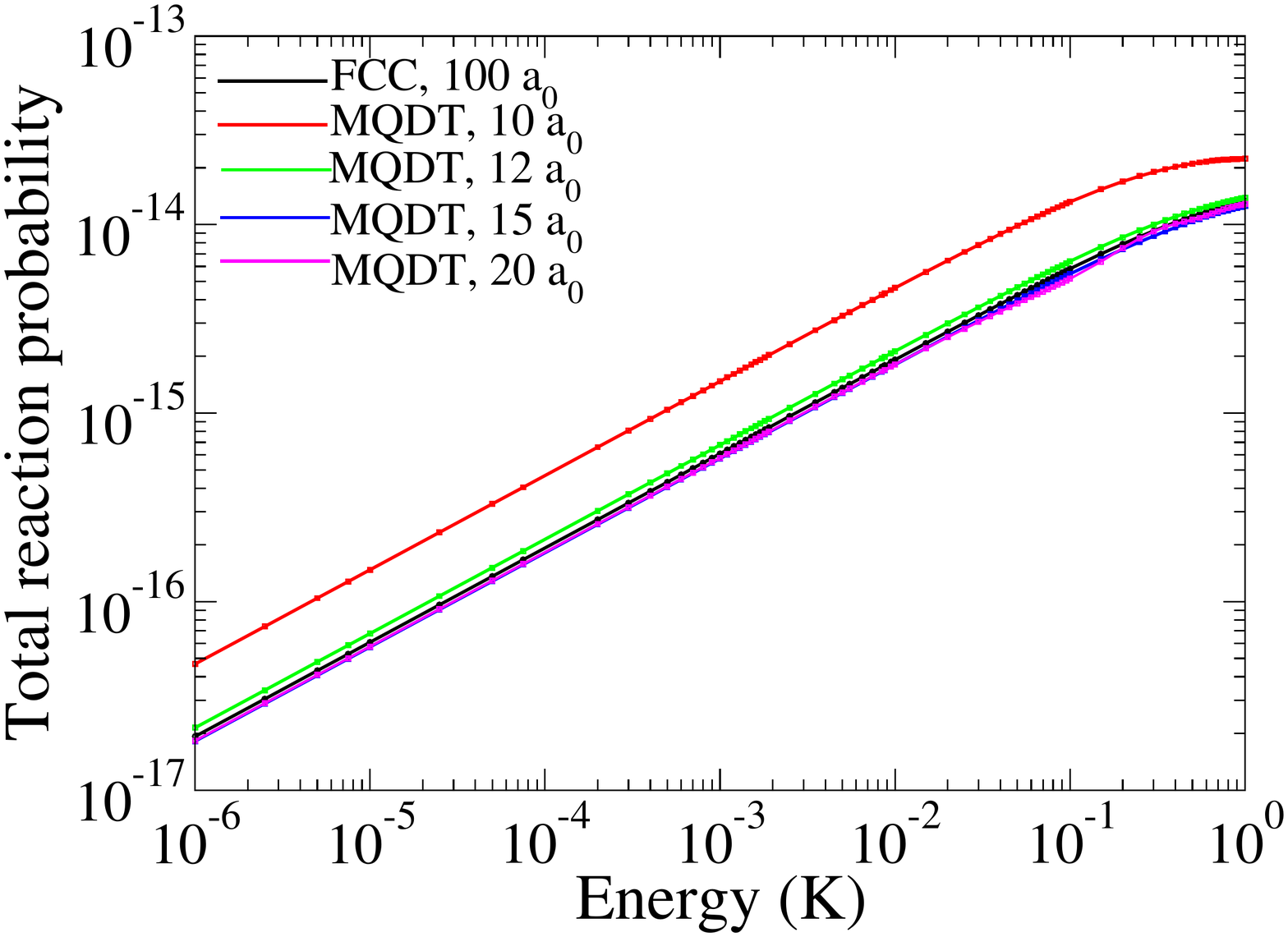}
\includegraphics[width=0.32\columnwidth]{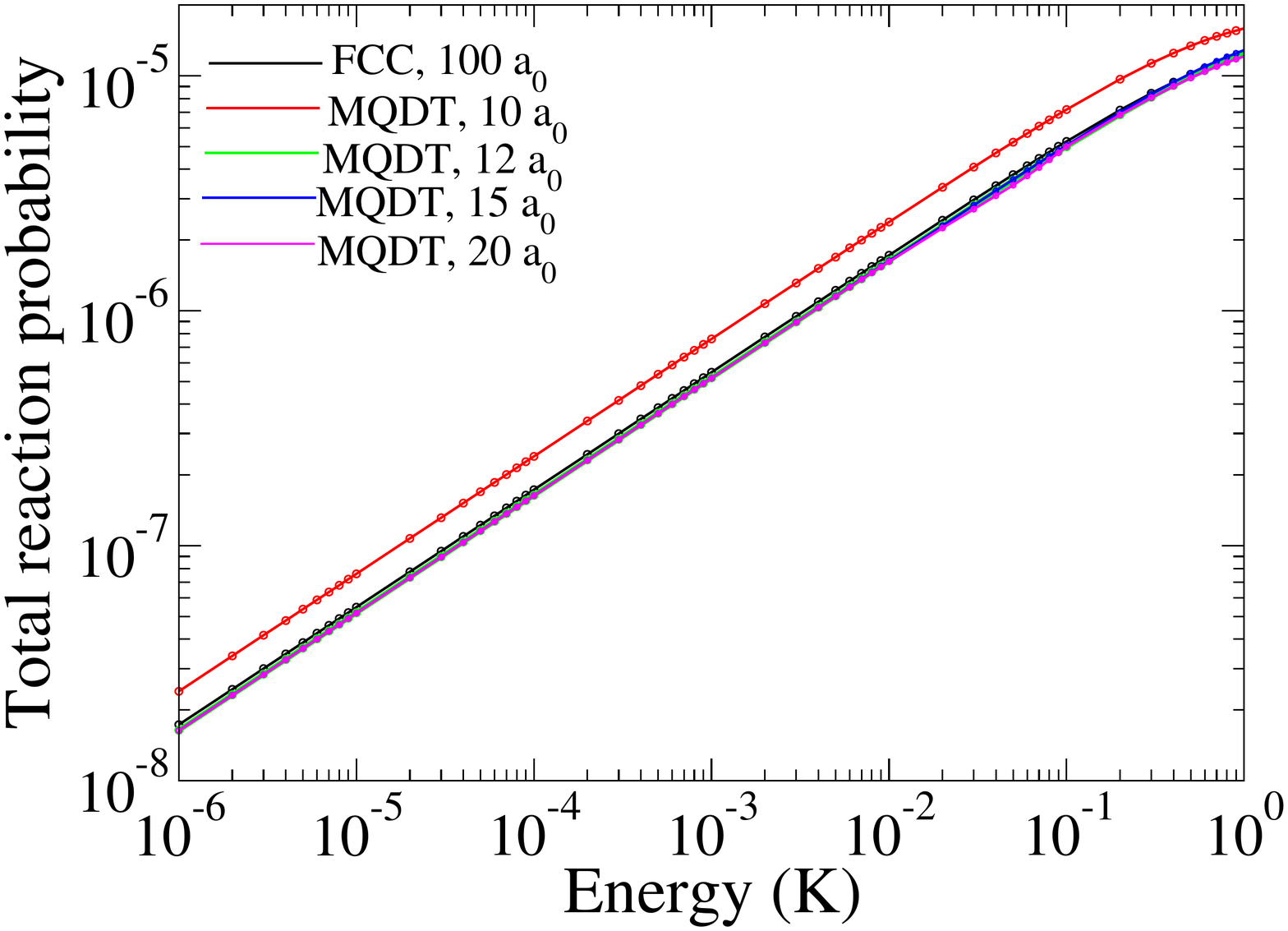}
\includegraphics[width=0.32\columnwidth]{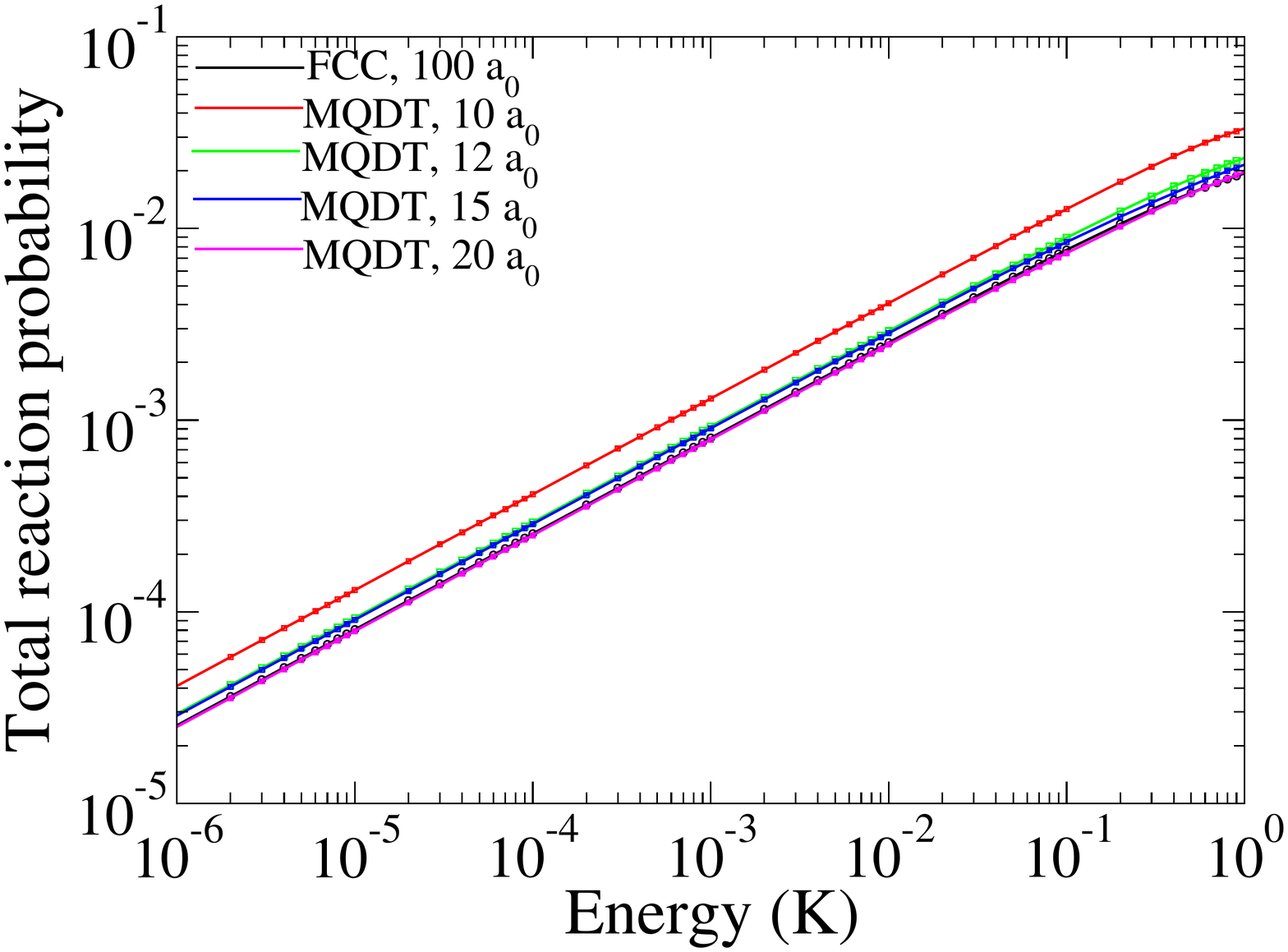}
\end{tabular}
\caption{(Color online) Convergence of the total reaction probability in the $1\mu$K-1 K regime for different values of 
the short-range matching distance that defines the boundary between  CC and MQDT formalisms. 
The left, middle and right panels correspond to $v=0,j=0$; $v=2,j=0$ and $v=5,j=0$ initial states of H$_2$, respectively.
The different curves in each panel correspond to different matching distances for the MQDT part:
10 $a_0$ (red curve), 12 $a_0$ (green curve), 15 $a_0$ (blue curve), and 20 $a_0$ (pink curve).}
\label{Convergence}
\end{figure}

\begin{figure}[tbp]
\centering
\includegraphics[width=0.8\columnwidth]{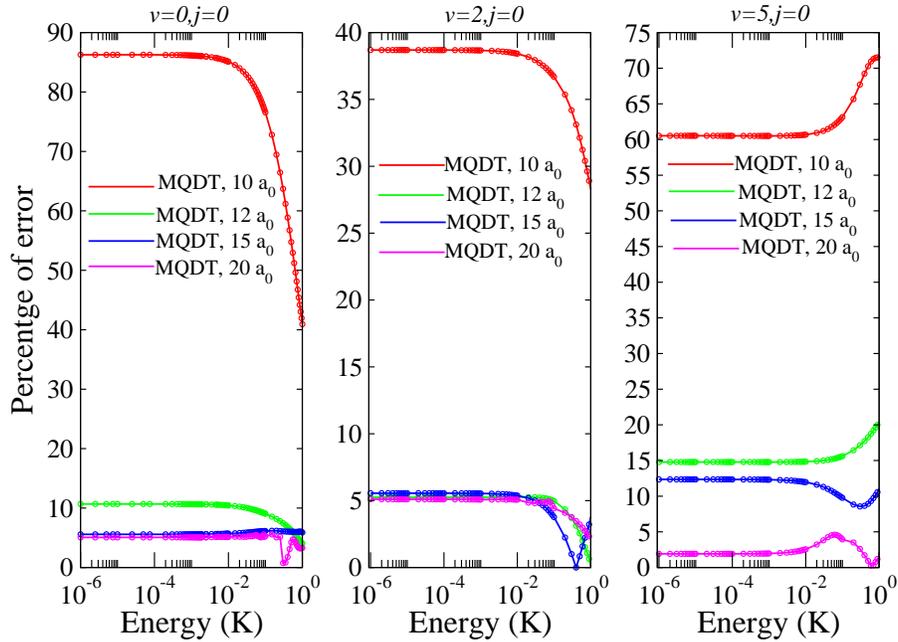}
\caption{(Color online) The percentage error in total reaction probability for different short-range matching radii $S_m$
as a function of the collision energy for the three different initial ro-vibrational states of H$_2$ depicted 
in Fig \ref{Convergence}.}
\label{error}
\end{figure}

Fig. \ref{Convergence} shows the convergence of the total reaction probabilities
as a function of the short-range matching distance $S_m$ for three different initial vibrational levels of the
H$_2$ molecule: $v=0,j=0$ (left panel); $v=2,j=0$ (middle panel) and $v=5,j=0$  
(right panel). In each of the panels, results are shown for four different values of $S_m$: 10, 12, 15, and
20 $a_0$. In all three panels, the black curve shows
converged full CC results obtained by matching at $\rho_{\infty} = 100 ~a_0$. The different colored curves correspond to  MQDT
results for different values of $S_m$. 
It is seen that the hybrid CC-MQDT method
yields nearly identical results as the numerically exact full CC calculation for a matching
distance of 20 $a_0$. The 
percentage errors for the different matching distances are presented in Fig.\ref{error} for 
the different initial vibrational levels. While ideally one would like to have the smallest value of $S_m$ possible, any 
choice of $S_m$ at which the interaction potential has not reached its true asymptotic form will lead to
larger errors as depicted by the results for $S_m$=10 $a_0$. Though any additional 
phase shift due to $V^{\rm lr}$ beyond $S_m$ is taken into account by MQDT, this phase shift is not taken into account for the reactive
and vibrational de-excitation channels that are characterized by high kinetic energies and not described by MQDT.
The percentage error 
is 30-80\% for a matching radius of 10 $a_0$ (red curves) for all three vibrational levels. The percentage error is less than  
10\% for $S_m=12$ $a_0$ (green curves)
and within 4-6\% for $S_m=15$ and 20 $a_0$ (blue and pink curves). These values are slightly higher for $v=5$ for the matching radii of 
12 and 15 $a_0$ but
less than 5\% for $S_m=20$ $a_0$. The convergence studies show that any values between 15-20 $a_0$ would be a reasonable short-range
matching distance for MQDT.

\begin{figure}[tbp]
\centering
\includegraphics[width=0.8\columnwidth]{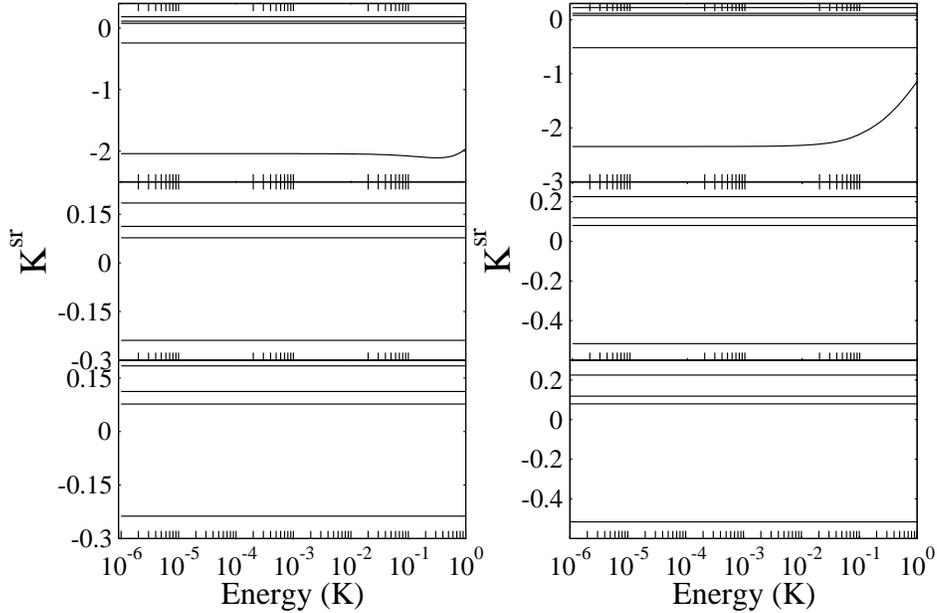}
\caption{Diagonal elements of ${\bf K}^{\rm sr}$ involving only the reactant block
as a function of energy for two different short-range matching distances. The left panel corresponds to 
a matching distance of 15 $a_0$ whereas the right panel pertains to the matching distance of 20 $a_0$.}
\label{shortkmat}
\end{figure}
\subsubsection{Energy Independence of ${\bf K}^{\rm sr}$:}
In Fig.\ref{shortkmat} we show the weak energy 
dependence of the diagonal elements of 
${\bf K}^{\rm sr}$ as a function of the kinetic energy  for two different matching distances of 15 (left panel) 
and 20 (right panel) $a_0$ 
for the $v=0,j=0$ initial state. 
It is clear from the left panel of Fig.\ref{shortkmat} that up to 100 mK the short range
${\bf K}$-matrix is independent of energy, but it becomes a smooth function of energy beyond 100 mK. This illustrates that, in principle,  
a single short-range ${\bf K}^{\rm sr}$-matrix can be used  in the 1$\mu$K-100 mK regime, even
though the cross sections vary by several orders of magnitude over this energy range.
For energies in the 100 mK-1 K regime an interpolation 
procedure over a sparse grid of energies may be used. Thus, for
the matching distance of 15 $a_0$, we divide the interpolation of $\bf{K^\text{sr}}$ into two different ranges of collision energy:
(i) the ultra-low energy range $E_c = 1 \mu K$ - 1 mK, where ${\bf K}^{\rm sr}$
is evaluated at $1\mu$K and 1 mK and (ii) the energy range $E_c =$ 100 mK - 1 K, where an
energy spacing of 200 mK is employed.

However, 
at $S_m=20$ $a_0$ (right panel of Fig.\ref{shortkmat}) a stronger dependence on energy is observed 
for the same elements of ${\bf K}^{\rm sr}$ beyond 10 mK. 
This appears to be due to the very shallow nature of the van der Waals potential well in the
D+H$_2$ system.
The minimum of the van der Waals potential for  D+H$_2$ is 
about 27 K at $S\approx 6.5$ $a_0$. As $S_m$ increases, the van der Waals well becomes shallower. At about
1 K the collision energy is no longer negligible compared to the well depth. Thus, the MQDT reference functions, 
defined by Eq.~(\ref{QDT_ref_f}) and (\ref{QDT_ref_g}), become energy sensitive. This is also reflected in the stronger energy dependence of the
${\bf K}^{\rm sr}$ matrix obtained at $S_m=20~a_0$.
Hence we need more points  
to accurately interpolate ${\bf K}^{\rm sr}$ evaluated at this matching distance. 
Therefore, in this case, we divide the interpolation of $\bf{K}^\text{sr}$ into three different ranges of collision energy: 
(i) the ultralow energy regime $E_c=1\mu$K - 1 mK, where ${\bf K}^{\rm sr}$ matrix is evaluated at $1\mu$K and 1 mK; (ii) the range $E_c$=10 mK-100 mK, where ${\bf K}^{\rm sr}$ is evaluated at 10 points with 10 mK separation; and
(iii) the range $E_c=$ 200 mK - 1 K, where an energy spacing of 100 mK is employed. The energy dependence of ${\bf K}^{\rm sr}$
indicates that when the scattering energies are only a
small fraction of the interaction potential (typically systems with deep attractive potential wells) a single ${\bf K}^{\rm sr}$ computed in the 
$\mu$K regime may
suffice to evaluate reaction cross sections in the Kelvin regime.
The results in Fig. \ref{error} and \ref{shortkmat} illustrate that 15 $a_0$ would be a resonable short-range matching radius for MQDT,
and we adopt this value for the rest of the calculations.

The ability of MQDT
to handle scattering in terms of an energy-smooth ${\bf K}^{\rm sr}$ will
carry over into more elaborate conditions than the ones presented here.
Notably, in the coldest collsions one may be interested
in the influence of electronic and nuclear spin states on scattering, to say nothing of the ability of a magnetic field
to manipulate scattering. This will require a description of Fano-Feshbach resonances, wherein the scattering observables
will vary widely on the resonant scale. In MQDT, a whole forest of resonances may be usefully
described over a large range of energy and field, utilizing a simple ${\bf K}^{\rm sr}$. Such resonances are not present in 
the prototype system we consider here, however.

\subsection{Initial state-selected reaction probabilities and cross sections}
\begin{figure}[tbp]
\centering
\includegraphics[width=1.0\columnwidth]{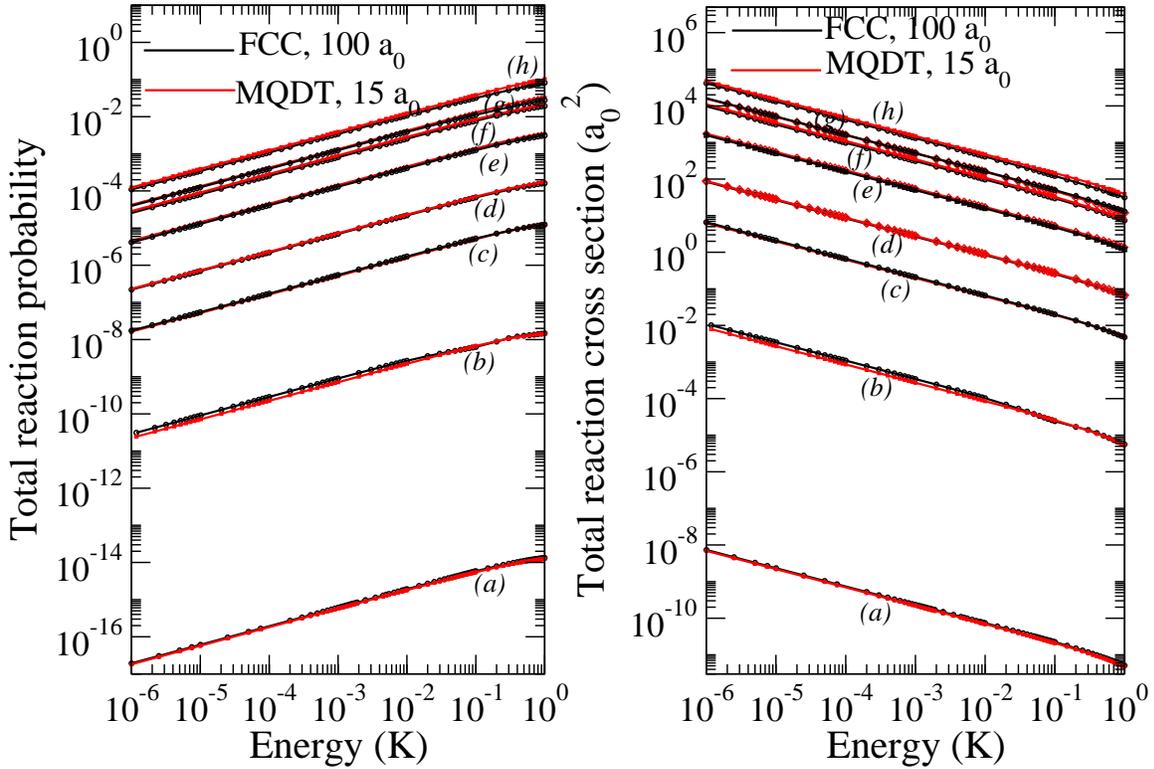}
\caption{Total reaction probabilities (left panel) and the corresponding  cross sections (right panel) for D+H$_2(v,j)\to$HD+H
reaction for $v=0-7$ and $j=0$ as functions of the incident collision energy. The
black curves denote full CC results, and the overlapping red curves denote the corresponding MQDT results. The different curves correspond to 
the different
initial vibrational levels of H$_2$: (a) $v=0$; (b) $v=1$; (c) $v=2$; (d) $v=3$;
(e) $v=4$; (f) $v=5$; (g) $v=6$; and (h) $v=7$.}
\label{Totalreaction}
\end{figure}

In Fig. \ref{Totalreaction} we present a comparison of total reaction probabilities and the corresponding cross sections  
for D+H$_2 (v=0-7,j=0)$ collisions evaluated using
the full CC and CC-MQDT approaches. Only the contribution from $J=0$ is included.
The full CC
results are obtained at an asymptotic matching distance of 100 $a_0$; whereas, the MQDT results use
a short-range matching distance of $S_m=15$ $a_0$.
The different curves (in both the left and right panels)
correspond to the different initial vibrational levels of the H$_2$ molecule. 
The solid black curves refer to results from full CC calculation, while the red curves correspond 
to MQDT results. The agreement between full CC and MQDT results
is excellent, and in most cases the percentage error is
about 5-6\%.
Since the reaction has an energy barrier of about 0.42 eV ($\sim$ 4800 K), the reaction probabilities are very small for 
$v=0$ and 1 initial states. However, the reactivity increases rapidly with H$_2$ vibrational excitation in agreement with previous results of 
Simbotin et al. \cite{Cote}. 
The reactivity for $v=6$ and 7 are comparable to that of $v=5$. This is an indication
that the reaction becomes 
essentially barrierless for $v\geq 5$.  It is  encouraging to see that, despite the  small reactivity for $v=0$ and 1, the 
MQDT approach is still able to quantitatively reproduce the full CC results. 

Regarding the
parameters employed in the full CC calculations, stepsize of $\Delta \rho = 0.01 a_0$
is adopted for the radial integration.
In the ABC code the total number of couple-channels is controlled by the three parameters $E_{\rm max}$, $j_{\rm max}$, and $\Omega_{\rm max}$.
For a given $E_{\rm max}$, all channels
with asymptotic rovibrational energies $\epsilon_{\tau vj}\leq E_{\rm max}$ are included, while the 
parameter $j_{\rm max}$ restricts the number of rotational levels for the different diatomic fragments. 
Similarly, $\Omega_{\rm max}$ ($k_{max}$ in the notation of ABC code) restricts the number of 
BF projection quantum numbers. Obviously, $\Omega_{max}=0$ for $J=0$.
We have chosen different values of $E_{\rm max}$ for three different ranges of initial vibrational levels of H$_2$:
for $v\leq 2$, $E_{\rm max} = 3.5$ 
eV (this leads to $v_{\rm max}$=7 for H$_2$ and $v_{\rm max}$=8 for HD in the basis sets); for $3\leq v\leq 6$, $E_{\rm max} = 4.25$ eV 
(this includes $v_{\rm max}$=10 for H$_2$ and $v_{\rm max}$=12 for HD) and for $v=7$, 
$E_{\rm max} = 4.75$ eV (this includes $v_{\rm max}$=14 for H$_2$ and $v_{\rm max}$=17 for HD).
However, we have restricted the total number of channels by fixing  $j_{\rm max} = 8$. Thus, the results presented
here are only partially converged with respect to the number of rotational levels included in the basis set. Fully converged full CC results 
using higher rotational levels in the basis 
set have already been reported by Simbotin et al. \cite{Cote} using the BKMP2 PES for the D+H$_2$ reaction for
the vibrational levels and energy regimes investigated in this study. Since the aim of this paper is to illustrate the usefulness of the
hybrid CC-MQDT formalism for cold and ultracold reactions, we resort to the smaller
basis set described above. 

\subsection{Rovibrational-state resolution of reaction products}
\begin{figure}[tbp]
\centering
\includegraphics[width=1.0\columnwidth]{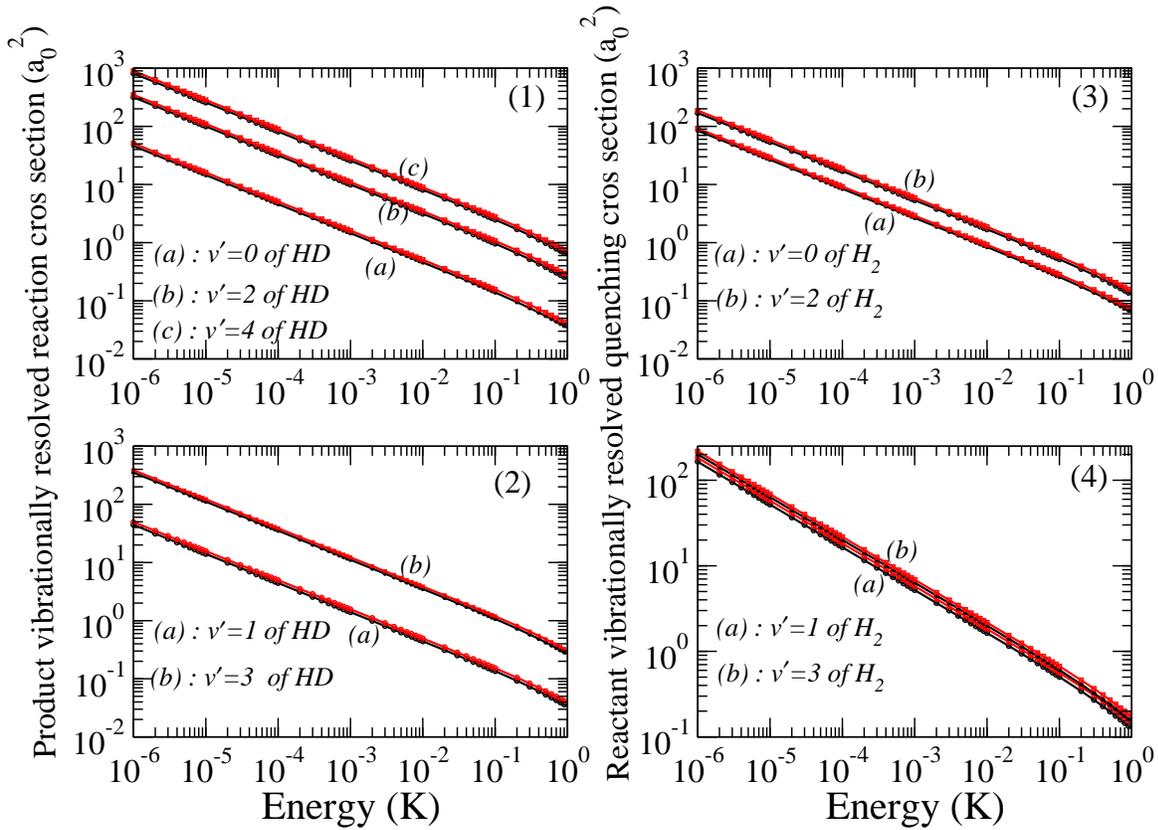}
\caption{(Color online) Comparison between full CC and MQDT calculations for vibrationally resolved HD product as well as non-reactively
scattered H$_2$ in 
D + H$_2 (v=4,j=0)$ collisions for total angular momentum $J=0$. 
The left panel shows vibrationally resolved cross sections for the HD product; whereas, the panel on the right shows 
corresponding cross sections for H$_2$ quenching.}
\label{vib}
\end{figure}
The CC-MQDT approach presented here allows full quantum state resolution of reaction products as in full CC
calculations, an aspect 
missing from previous MQDT
treatments of ultracold chemistry. As an illustrative example
we choose to study the $v=4,j=0$ initial state of H$_2$, which allows for the population of several rovibrational levels of 
the HD molecule as well as non-reactive quenching, leading to the population of lower vibrational levels of the reactant
molecule.
 The total reaction probability and cross section for $J=0$ for this initial H$_2$ level have already been shown in 
 Fig.\ref{Totalreaction}. In the left panels (1 and 2) of Fig.\ref{vib}, we show the corresponding vibrational-level resolved 
reaction cross sections for the HD product. Cross sections for non-reactive vibrational quenching of the H$_2$ molecule are shown in the 
right panels (3 and 4) of Fig.\ref{vib}. In all of the panels,
the black curves denote the full CC results, and the red curves depict the MQDT results. The 
agreement is excellent, reflecting the similar agreement for the total cross sections.

The different curves correspond to the different vibrational levels of the HD product or non-reactively
scattered H$_2$ molecule. Vibrational levels $v'=0-4$ of the HD molecule are populated in the reaction. 
They are depicted by curves $(a)-(c)$ in panel (1) and $(a)$ and $(b)$ in panel (2) of Fig.\ref{vib}.
The curves labeled $(a)$ and $(b)$ in panels (3) and (4) of Fig.\ref{vib} show H$_2$ quenching cross sections.
The results show that, even in the 
ultracold limit, chemical reaction dominates over inelastic vibrational quenching. 
The vibrationally resolved cross sections for quenching and reaction are proportional to the following sums: 
$\sigma^{v'}_{ \tau vj\Omega,\rm qn} \propto \sum_{{\rm qn},\hspace{0.1cm} j'\Omega'}|
S^{v'}_{\tau vj\Omega\to \tau v'j'\Omega'}(E_c)|^2$
and $\sigma^{v'}_{ \tau vj\Omega,\rm re} \propto \sum_{{\rm re},\hspace{0.1cm} j'\Omega'}|
S^{v'}_{\tau vj\Omega\to \tau'v'j'\Omega'}(E_c)|^2$, respectively,
where the sums are carried over all open rotational levels within a given vibrational level, $v'$, of a particular arrangement channel. 
Note that here we omit
the quantum number $J$ to simplify the notation.  

\begin{figure}[htbp]
\centering
\includegraphics[width=1.0\columnwidth]{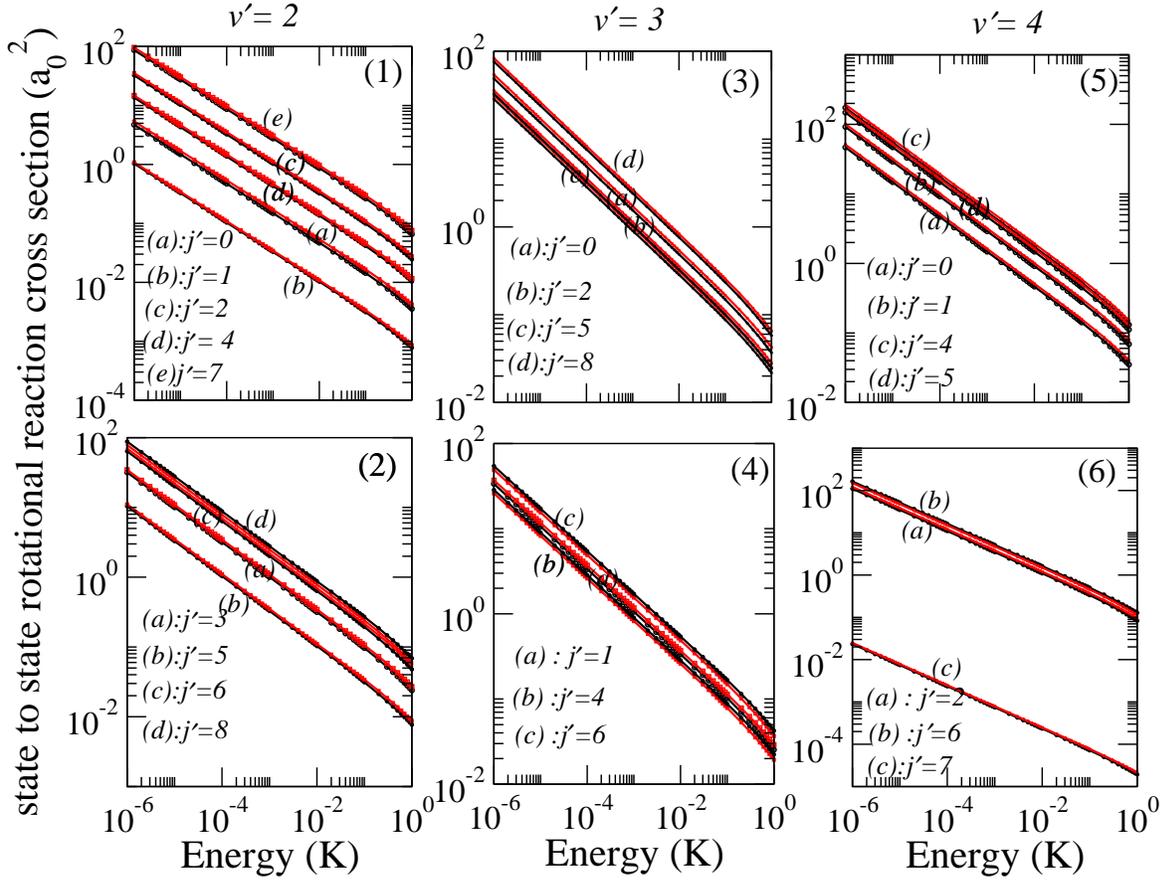}
\caption{(Color online) Comparison between full CC and MQDT calculations for rotationally resolved
reaction cross sections for the HD product in D + H$_2 (v=4,j=0)$ collisions. The left panels (1 and 2) correspond to
HD$(v'=2,j')$, the middle panels (3 and 4) correspond to HD$(v'=3,j')$, and
the right panels (5 and 6) correspond to HD$(v'=4,j')$. }
\label{rot_reac}
\end{figure}
\begin{figure}[tbp]
\centering
\includegraphics[width=1.0\columnwidth]{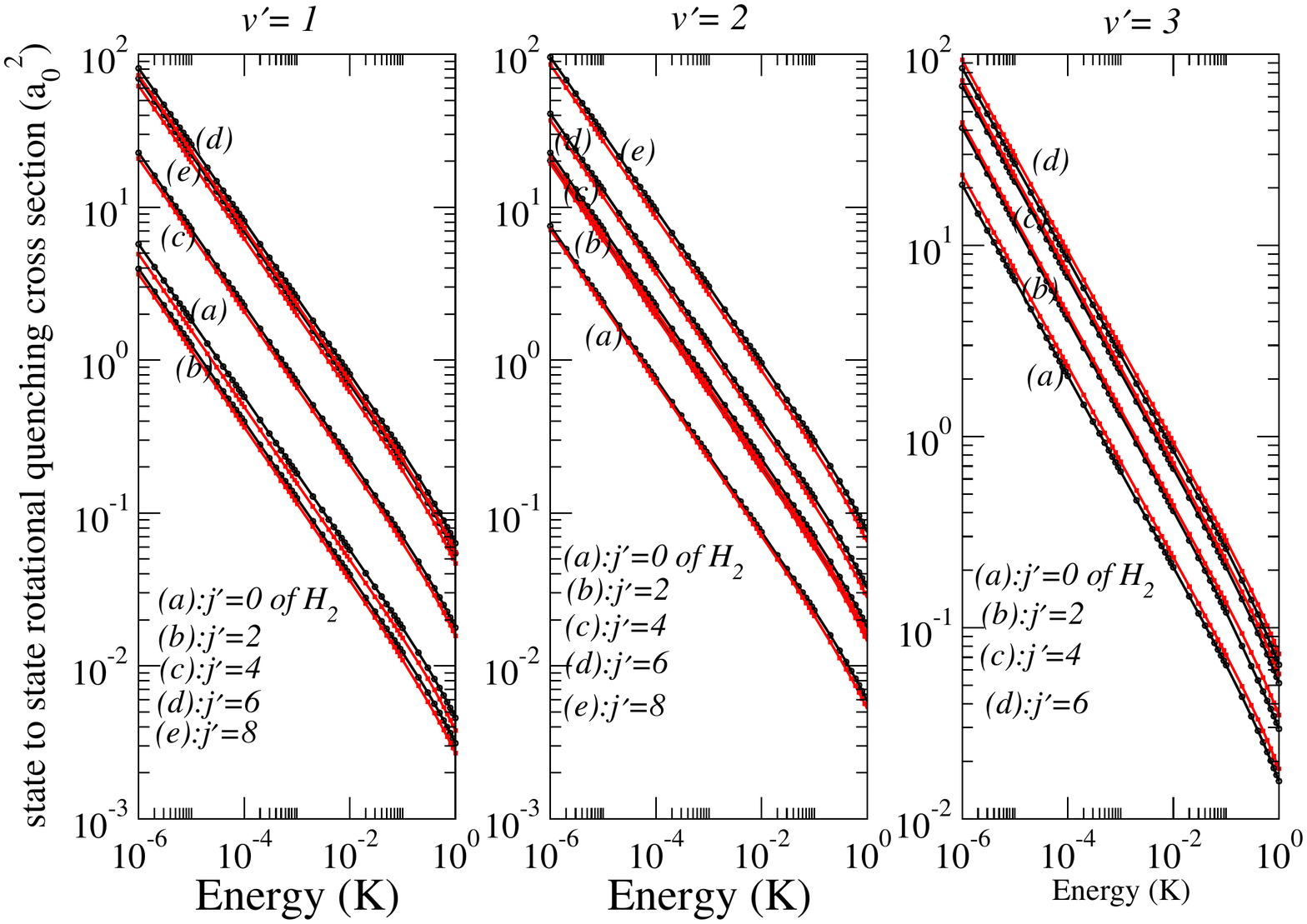}
\caption{(Color Online) Similar results as in Fig.\ref{rot_reac} but for the non-reactively scattered H$_2$.
 The left panel corresponds to
H$_2(v'=1,j')$, the middle panel corresponds to H$_2(v'=2,j')$, and
the right panel corresponds to H$_2(v'=3,j')$.}
\label{rot_quen}
\end{figure}

For vibrationally excited molecules, the treatment of MQDT is slightly modified in addition 
to the procedure described in Sec. IIB. For these cases, the MQDT-formalism is only applied
to the initial collisional channel and those associated with cold and ultracold energies. Whereas, the other open inelastic
(non-reactive) channels characterized by high kinetic energies
within the reactant arrangement are treated in the standard way, i.e., 
these channels are matched to the normal asymptotic Bessel functions. In other words, they are treated like product channels. Thus, 
the transformations described by Eqs. (\ref{trans1}) - (\ref{trans4}) are also applied to the high 
kinetic energy channels of the reactant block in ${\bf \tilde K}_{\rm RR}$.

\begin{figure}[htbp]
\centering
\includegraphics[width=0.8\columnwidth]{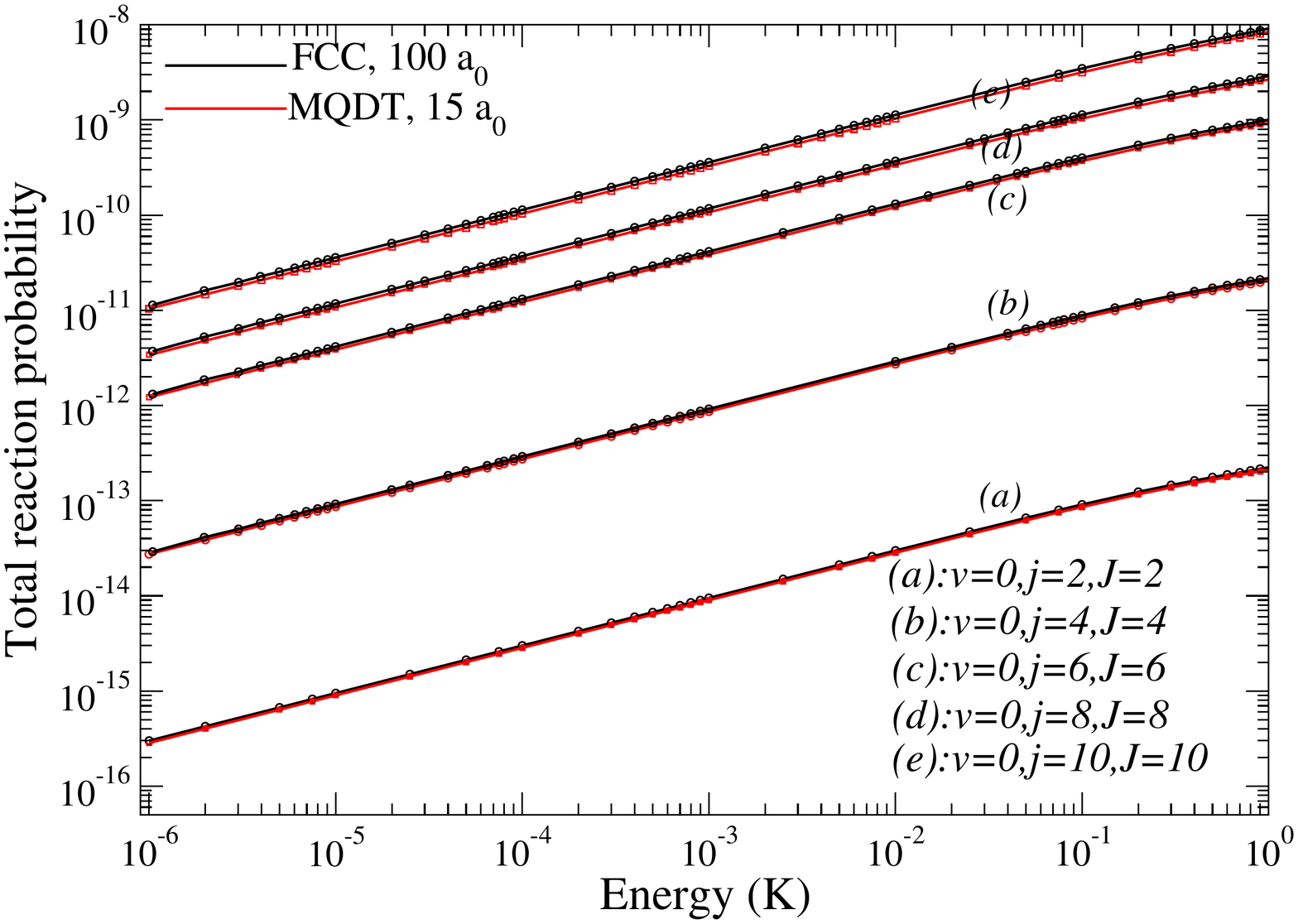}
\caption{(Color online) Comparision of total reaction probability between the full CC and MQDT calculations as a function of collisional 
energy $E_c$ 
for an initially rotationally excited H$_2$ molecule within the $v=0$ vibrational level.}
\label{NonJ}
\end{figure}

A comparison between the full CC and MQDT methods for rotationally resolved cross sections for the HD product in different 
energetically open vibrational 
levels 
are presented for the $v=4$ initial state in Fig. \ref{rot_reac}. The different panels correspond to the rotational 
distributions in the three highest populated vibrational levels, $v'=2,3,$ and 4 of HD. As in other cases, the black curves denote the 
full CC results;
whereas, the red curves depict the  MQDT calculation. Again,  
the agreement is excellent, as it is 
in the case of vibrational distribution.  Any small deviation can be attributed, 
at least in part, to neglecting the anisotropic contribution to the interaction potential in the construction of the MQDT reference functions.

The rotational level populations of non-reactively scattered H$_2$ molecules in three highest populated vibrational levels ($v'=1,~2$ and 3)  
are presented 
in Fig.\ref{rot_quen}. As before, the full CC results are shown by the black curves, and the MQDT-results are shown by the red curves.
Note that only even rotational levels are populated since the calculations are restricted to the even parity states 
of the H$_2$ molecule 
(para-H$_2$).
\subsection{Total reactive probability for non-zero $J$}
Finally, in Fig.\ref{NonJ} we present a comparison between full CC and MQDT results for the total reaction probability for
different initial rotational levels of H$_2$ within the $v=0$ vibrational level. The different curves
labeled (a)-(e) correspond to 
the initial rotational levels $j=2,~4,~6, 8$ and 10, respectively. For each case, the calculations are restricted to $j=J$ to capture $s$-wave
scattering in the incident channel. Also,
for this particular case, both the triatomic parity $p$ and diatomic parity $q$
are limited to +1. These results are included to demonstrate that the method is not restricted to the 
non-rotating case.

\section{Conclusions}
\label{section:conclusions}
We have presented a formulation of multichannel quantum defect theory that is able to yield full rovibrational level resolved 
cross sections and rate coefficients for cold and ultracold  chemical reactions with an accuracy comparable to the
full close-coupling calculations but at a much reduced computational cost. The
method makes use of the close-coupling approach but restricts it to the chemically relevant region with the 
long-range part handled by the MQDT formalism. The usefulness and robustness of the method is  
illustrated by applying it to the benchmark D+H$_2\to$ HD+H reaction for vibrational levels $v=0-7$ of the
H$_2$ molecule. Rotational and vibrational populations of the product HD molecule are evaluated using
this hybrid CC-MQDT method, and they are shown to be comparable to those obtained from the full close-coupling calculations. A similar
agreement is found for rovibrational distributions of the non-reactively scattered H$_2$ molecule.

The method has many attractive features that make it appealing for ultracold chemical reactions. For instance, the 
short-range ${\bf K}$-matrix, evaluated by matching MQDT reference functions to the log-derivative matrix from
the close-coupling calculation, is found to be largely energy independent in the $1\mu$K-10 mK regime. This implies 
that CC calculations need to be performed just at a single collision energy, say $1\mu$K, to evaluate cross sections 
at all energies between $1\mu$K-10 mK. Since this is the time consuming part of the computation, it
 leads to significant savings in computational time. For systems with deeper interaction potentials 
than D+H$_2$ (which has repulsive interaction at short range and a shallow van der Waals well in the long-range),
the energy independence of the short-range ${\bf K}$-matrix is expected to be valid over a larger range of energies, 
allowing one to restrict the CC calculations to a few collision energies in the ultracold to 1-10 K regime. 
One caveat remains, namely, that the short-range ${\bf K}$-matrix may contain resonances from ro-vibrational channels that 
become closed at radial distances smaller than $S_m$ \cite{Hud1}, potentially requiring special handling within MQDT. 
Such resonances did not occur in the present calculation, however. Similarly,
for open-shell systems with deep potential wells, calculations involving external fields, spin, and hyperfine effects can all 
be restricted to the MQDT part as energy splitting due to these factors will be many orders of magnitude smaller
than the well depth of the interaction potential at short-range.
 
The application of the method to more complex reactions with deeper potential wells and inclusion 
of external field effects are planned.

\section{Acknowledgements}
This work was supported  in part by NSF grant  PHY-1205838 (N.B.) and
ARO MURI grant No. W911NF-12-1-0476 (N.B. and J.L.B).
JH is grateful to Brian Kendrick for many helpful discussions.

\renewcommand{\theequation}{A-\arabic{equation}}
\setcounter{equation}{0}  
\section*{APPENDIX-A}  
The  matrix elements of the overlap matrix ${\bf O}$ between 
different arrangements $(\tau,\tau') $ can be expressed in the SF representation as \cite{Krems}
\begin{eqnarray}
O_{\tau vj\ell,\tau'v'j'\ell'} &=& \frac{4}{\sin 2\theta_{\tau}\sin 2\theta_{\tau'}} \langle \Upsilon_{\tau vj}
(\theta_{\tau};\rho)|\langle{\cal Y}^{JM}_{j\ell}({\hat s}_{\tau},{\hat S}_{\tau})| \Upsilon_{\tau' v'j'}
(\theta_{\tau'};\rho)\rangle|{\cal Y}^{JM}_{j'\ell'}({\hat s}_{\tau'},{\hat S}_{\tau'})\rangle \nonumber \\
&=& \int d{\hat S_{\tau}}d{\hat s_{\tau}} \int^{\pi/2}_{0} d\theta_{\tau}
{\cal Y}^{JM}_{j\ell}({\hat s}_{\tau},{\hat S}_{\tau}){\cal Y}^{JM}_{j'\ell'}({\hat s}_{\tau'},{\hat S}_{\tau'}) 
\nonumber \\ &\times& \left[\frac{\sin 2\theta_{\tau}}
{\sin 2\theta_{\tau'}}\right]\Upsilon_{\tau vj}
(\theta_{\tau};\rho) \Upsilon_{\tau'v'j'}
(\theta_{\tau'};\rho), \label{overlapmat}
\end{eqnarray}
where the Jacobian, $\frac{1}{4}\sin^2 2\theta_{\tau}$, for the integration over the angles is included in arriving at the second expression. 
The orthonormality of the matrix elements of ${\bf O}$ within the same arrangement,
$(\tau,\tau)$ yields ${\bf O}_{\tau vj\ell,\tau v'j'\ell'} = \delta_{vv'}\delta_{jj'}\delta_{\ell,\ell'}$.

The matrix elements in the SF representation of the adibatic Hamiltonian are given by 
\begin{eqnarray}
\langle \xi_{m}|H_{\rm ad}(\rho)|\xi_{m'}\rangle &= &\frac{4}{\sin 2\theta_{\tau}\sin 2\theta_{\tau'}}\frac{1}{\sqrt{\sigma_{m} \sigma_{m'}}}
\sum_{\tau v j\ell}\sum_{\tau'v'j'\ell'}X_{\tau v j\ell,m}(\rho)X_{\tau'v'j'\ell',m'}(\rho)\nonumber 
\\ &\times&\langle \Upsilon_{\tau vj}
(\theta_{\tau};\rho)|\langle{\cal Y}^{JM}_{j\ell}({\hat s}_{\tau},{\hat S}_{\tau})|H_{\rm ad}(\rho) |
\nonumber \\ &\times&\Upsilon_{\tau' v'j'}
(\theta_{\tau'};\rho)\rangle|{\cal Y}^{JM}_{j'\ell'}({\hat s}_{\tau'},{\hat S}_{\tau'})\rangle \label{MatAdia}.
\end{eqnarray}
In the SF representation, the sector-to-sector transformation matrix elements are defined by the following expression
\begin{eqnarray}
[{\bf S}(\rho_{j},\rho_{j+1})]_{nn'} &=& \langle \Phi^{JM}_{n}(\omega;\rho_j)|\Phi^{JM}_{n'}(\omega;\rho_{j+1})\rangle \nonumber \\ 
&=& \sum_{m,m'}F_{mn}(\rho_j)F_{m'n'}(\rho_{j+1}) \frac{1}{\sqrt{\sigma_{m}\sigma_{m'}}}\sum_{\tau vj\ell}\sum_{\tau'v'j'\ell'}
X_{\tau v j\ell,m}(\rho_j)X_{\tau'v'j'\ell',m'}(\rho_{j+1})
\nonumber \\&\times&\frac{4}{\sin 2\theta_{\tau}\sin 2\theta_{\tau'}}\langle \Upsilon_{\tau vj}
(\theta_{\tau};\rho_j)|\langle {\cal Y}^{JM}_{j\ell}({\hat s}_{\tau},{\hat S}_{\tau})|\nonumber \\&\times&\Upsilon_{\tau' v'j'}
(\theta_{\tau'};\rho_{j+1})\rangle|{\cal Y}^{JM}_{j'\ell'}({\hat s}_{\tau'},{\hat S}_{\tau'})\rangle.\label{sectoroverlap}
\end{eqnarray}

\renewcommand{\theequation}{C-\arabic{equation}}
\setcounter{equation}{0} 
\section*{APPENDIX-B}
\subsection*{Scattering Boundary Conditions}
Once the propagation of the logderivative matrix from a small $\rho$  to an asymptotically
 large value of $\rho$ is accomplished, one needs to apply the asymptotic boundary conditions to the 
${\bf Y}$ matrix (calculated in DC) to obtain ${\bf K}^J$. The boundary conditions are applied in Jacobi coordinates since the asymptotic 
forms of the 
wavefunctions are well known in this coordinate. Therefore, asymptotic analysis 
involves projecting the DC wavefunctions onto wavefunctions in Jacobi coordinates. 
A detailed description of this procedure is given by Pack and Parker~\cite{Pack} and only  brief 
account is given below. Note that a similar approach will be used for matching MQDT reference functions to 
the log-derivative matrix from the CC calculations. 
First, in the asymptotic limit ($\rho\longrightarrow \infty$), where the exchange interactions between different arrangements become zero,
it is convenient to express the total wavefunction $\Psi^{JM}$ of the reactive scattering system in  Jacobi coordinates:
\begin{equation}
\Psi^{JM} = \sum_{\tau v j \ell}\frac{1}{s_{\tau}S_{\tau}}G^J_{\tau v j \ell}(S_{\tau}){\cal X}_{\tau v j}(s_{\tau})
{\cal Y}^{JM}_{j\ell}({\hat s}_{\tau},{\hat S}_{\tau}),\label{asymptoJC}
\end{equation}
where the quantities $G^J_{\tau v j \ell}$ and ${\cal X}_{\tau v j}$ denote, respectively, the radial expansion  coefficients and the vibrational wavefunctions of the
diatomic molecule. Note that here we explicitly include the quantum number $n=\{\tau, v,j,\ell\}$ since it refers to the asymptotic molecuar states.
In the asymptotic region,  the angular
functions ${\cal Y}^{JM}_{j\ell}({\hat s}_{\tau},{\hat S}_{\tau})$ are the same in both Jacobi and DC as there is no overlap between functions 
with different $\tau$. 
The ro-vibrational wavefunctions  
in both coordinate systems satisfy the
orthonormal condition. The asymptotic form of the total wavefunction in DC is given by
\begin{equation}
\Psi^{JM} = 2\sum_{\tau v j \ell}\frac{\Gamma^J_{\tau v j \ell}(\rho)}{\rho^{5/2}}\frac{\Upsilon_{\tau v j}
(\theta_{\tau};\rho)}{\sin 2\theta_{\tau}}
{\cal Y}^{JM}_{j\ell}({\hat s}_{\tau},{\hat S}_{\tau}).
\end{equation}
By usual projection and using the orthonormality of the ro-vibrational wave functions,  the following expressions for the radial expansion 
coefficients, 
$\Gamma_{\tau v j \ell}^J(\rho)$ and its derivative, $\frac{\partial \Gamma_{\tau v j \ell}^J(\rho)}{\partial \rho}$ are obtained \cite{Pack}
\begin{eqnarray}
\Gamma_{\tau v j \ell}^J(\rho) &=& \frac{1}{4}\int_0^{\pi/2}d\theta_{\tau'}(\sin 2\theta_{\tau'})^2 \int d{\hat s}_{\tau'}d{\hat S}_{\tau'}
 \nonumber \\ &\times&\left[\frac{2\Upsilon_{\tau' v' j'}^*(\theta_{\tau'};\rho) {\cal Y}^{JM*}_{j'\ell'}({\hat s}_{\tau'},{\hat S}_{\tau'})}
 {\sin 2\theta_{\tau'}}\right]\left(\Psi^{JM}\rho^{5/2}\right)\label{proj1}\\
 \frac{\partial \Gamma_{\tau v j \ell}^J(\rho)}{\partial \rho}&=&\frac{1}{4}\int_0^{\pi/2}d\theta_{\tau'}(\sin 2\theta_{\tau'})^2 \int
 d{\hat s}_{\tau'}d{\hat S}_{\tau'}
 \nonumber \\ &\times&\left[\frac{2\Upsilon_{\tau'v'j'}^*(\theta_{\tau'};\rho) {\cal Y}^{JM*}_{j'\ell'}({\hat s}_{\tau'},{\hat S}_{\tau'})}
 {\sin 2\theta_{\tau'}}\right]\frac{\partial}{\partial \rho}\left(\Psi^{JM}\rho^{5/2}\right).\label{proj2}
\end{eqnarray}
Using $sS = \rho^2\sin\theta\cos\theta=\frac{\rho^2\sin 2\theta}{2}$ and multiplying by $\rho^{5/2}$, Eq.(\ref{asymptoJC}) is reexpressed as
\begin{eqnarray}
 \rho^{5/2}\Psi^{JM} = \sum_{\tau v j \ell}\frac{2\rho^{1/2}}{\sin 2\theta_{\tau}}G^J_{\tau v j \ell}(S_{\tau}){\cal X}_{\tau v j}(s_{\tau})
{\cal Y}^{JM}_{j\ell}({\hat s}_{\tau},{\hat S}_{\tau}).
\end{eqnarray}
Substitution of the above equation in Eqs. (\ref{proj1}) and (\ref{proj2}) and performing the integrals over $d{\hat s}_{\tau'}$ 
and $d{\hat S}_{\tau'}$ and using
$\int d{\hat s}_{\tau'} d{\hat S}_{\tau'}  {\cal Y}^{JM*}_{j'\ell'}({\hat s}_{\tau'},{\hat S}_{\tau'}){\cal Y}^{JM}_{j\ell}
({\hat s}_{\tau},{\hat S}_{\tau})=
\delta_{\tau'\tau}\delta_{j'j}\delta_{\ell'\ell}$, one 
obtains
 \begin{eqnarray}
\Gamma_{\tau v j \ell}^J(\rho) &=&\sum_{\tau v j \ell}\delta_{\tau'\tau}\delta_{j'j}\delta_{\ell'\ell}
\rho^{1/2}\int_0^{\pi/2}d\theta_{\tau'}\Upsilon_{\tau',v',j'}^*(\theta_{\tau'};\rho)G^J_{\tau v j \ell}(S_{\tau}){\cal X}_{\tau v j}(s_{\tau})\\
 \frac{\partial \Gamma_{\tau v j \ell}^J(\rho)}{\partial \rho}&=&\frac{1}{2\rho}\Gamma_{\tau v j \ell}^J(\rho)+\sum_{\tau v j \ell} 
 \delta_{\tau'\tau}\delta_{j'j}\delta_{\ell'\ell}\rho^{1/2}\nonumber 
 \\ &\times& \int_0^{\pi/2}
 d\theta_{\tau'}\Upsilon_{\tau' v j}^*(\theta_{\tau'};\rho)\frac{\partial[G^J_{\tau v j \ell}(S_{\tau}){\cal X}_{\tau v j}(s_{\tau})]}{\partial\rho}.
\end{eqnarray}
The asymptotic boundary conditions of the wavefunction and its derivative in Jacobi coordinates, in terms of two reference 
functions ${\bf \hat f}$ and ${\bf \hat g}$, are ${\bf G}(S_{\tau}) = {\bf \hat f}(S_{\tau})-{\bf \hat g}(S_{\tau})\bf K$ and 
${\bf G'}(S_{\tau}) = {\bf \hat f'}(S_{\tau})-{\bf \hat g'}(S_{\tau})\bf K$, respectively,
where the primes indicate  differentiation with respect to $S$, and $\bf K$ is the reactance matrix.
In our case, the reference functions ${\bf \hat f}$ and ${\bf \hat g}$ are spherical Bessel functions
in the product channels and MQDT-reference functions in the reactant
channels. These two boundary conditions
for the wavefunction $G$ in DC yields
\begin{eqnarray}
 {\bf \Gamma}^J(\rho) = {\bf {\cal A}}(\rho)-{\bf {\cal B}}(\rho){\bf K}^J\nonumber \\
 \frac{\partial {\bf \Gamma}^J(\rho)}{\partial \rho} = \frac{1}{2\rho}{\bf {\cal A}}(\rho)+{\bf {\cal C}}(\rho)-{\bf {\cal D}}(\rho){\bf K}^J,
\end{eqnarray}
where the matrix elements of  $\bf {\cal A}$ and $\bf {\cal C}$ have the form of
\begin{eqnarray}
 {\cal A}_{\tau' v' j' \ell',\tau v j \ell}(\rho) &=& \delta_{\tau'\tau}\delta_{j'j}\delta_{\ell'\ell}
 \rho^{1/2}\int_0^{\pi/2}d\theta_{\tau'}
 \Upsilon_{\tau' v' j'}^*(\theta_{\tau'};\rho) {\hat f}_{\tau v j \ell}(S_{\tau}){\cal X}_{\tau v j}(s_{\tau})\\
{\cal C}_{\tau' v' j' \ell',\tau v j \ell}(\rho) &=& \delta_{\tau'\tau}\delta_{j'j}\delta_{\ell'\ell}\rho^{1/2}
\int_0^{\pi/2}d\theta_{\tau'}\Upsilon_{\tau' v' j'}^*(\theta_{\tau'};\rho)\nonumber \\ &\times&\left[\frac{d {\hat f}_{\tau v j \ell}}
{dS_{\tau}}{\cal X}_{\tau v j}(s_{\tau})\cos\theta_{\tau}+
\frac{d {\cal X}_{\tau v j}(s_{\tau})}{ds_{\tau}}{\hat f}_{\tau v j \ell}\sin\theta_{\tau}\right].
\end{eqnarray}
The matrices ${\bf {\cal B}}$ and ${\bf {\cal D}}$ have similar expressions but with
the Bessel function ${\bf \hat f}(S)$ replaced by  ${\bf \hat  g}(S)$. It is clear from the expression of the projection matrices 
$ {\bf {\cal A}}$ and
$ {\bf {\cal C}}$ that they are diagonal in all the quantum numbers but there is an overlap between different vibrational states with same 
$\{\tau,j,\ell\}$. From Eq.(24) and using the above expressions, the final form of 
the K-matrix is obtained in terms of the log-derivative matrix (in DC),
\begin{equation}
 {\bf K}^J = \left({\bf Y {\cal B}}-\frac{1}{2\rho}{\bf  {\cal B}}-{\bf {\cal D}}\right)^{-1} 
 \left({\bf Y {\cal A}}-\frac{1}{2\rho}{\bf  {\cal A}}-{\bf {\cal C}}\right). \label{Kmat}
\end{equation}
The S-matrix, ${\bf S}^J$ is obtained from the K-matrix using the well known Cayley transformation.
It is to be noted that both $ {\bf K}^J$ and ${\bf S}^J$ are independent of $\rho$ and the choice of the coordinate systems.

\end{document}